\documentclass[manuscript]{aastex62}
\received{}
\revised{}
\accepted{}

\shorttitle{TTV search in the TrES-3 and Qatar-1 system  using TESS data}
\shortauthors{Mannaday et al.}

\usepackage{longtable}
\usepackage{floatrow}
\usepackage[utf8x]{inputenc}
\usepackage{xcolor}
\usepackage{multirow}
\begin{document}

\title{Revisiting the Transit Timing Variations in the TrES-3 and Qatar-1 systems with TESS data}

\correspondingauthor{Parijat Thakur}
\email{parijat@associates.iucaa.in, parijatthakur@yahoo.com}
\author{Vineet Kumar Mannaday}
\affil{Department of Pure and Applied Physics, Guru Ghasidas Vishwavidyalaya (A Central University), Bilaspur (C.G.) 495 009, India}
\affil{Department of Physics, Govt. Niranjan Kesharwani College, Kota, Bilaspur (C.G.) 495113, India}
\author[0000-0002-0786-7307]{Parijat Thakur}
\affiliation{Department of Pure and Applied Physics, Guru Ghasidas Vishwavidyalaya (A Central University), Bilaspur (C.G.) 495 009, India}
\author{John Southworth}
\affiliation{Astrophysics Group, Keele University, Staffordshire ST5 5BG, UK}
\author{Ing-Guey Jiang}
\affiliation{Department of Physics and Institute of Astronomy, National Tsing-Hua University, Hsinchu, Taiwan}
\author{D. K. Sahu}
\affiliation{Indian Institute of Astrophysics, Bangalore 560 034, India}
\author{Luigi Mancini}
\affiliation{Max Planck Institute for Astronomy, Königstuhl 17, D-69117 Heidelberg, Germany}
\author{M. Va\v{n}ko}
\affiliation{Astronomical Institute, Slovak Academy of Sciences, SK-059 60 Tatransk\'{a} Lomnica, Slovakia}
\author{Emil Kundra}
\affiliation{Astronomical Institute, Slovak Academy of Sciences, SK-059 60 Tatransk\'{a} Lomnica, Slovakia}
\author{Pavol Gajdo\v{s}}
\affiliation{Institute of Physics, Faculty of Science, Pavol Jozef \v{S}af\'{a}rik University, 040 01 Ko\v{s}ice, Slovakia}
\author{Napaporn A-thano}
\affiliation{Department of Physics and Institute of Astronomy, National Tsing-Hua University, Hsinchu, Taiwan}
\author{Devesh P. Sariya}
\affiliation{Department of Physics and Institute of Astronomy, National Tsing-Hua University, Hsinchu, Taiwan}
\author{Li-Chin Yeh}
\affiliation{Institute of Computational and Modeling Science, National Tsing-Hua University, Hsinchu 30013, Taiwan}
\author{Evgeny Griv}
\affiliation{Department of Physics, Ben-Gurion University, Beer-Sheva 84105, Israel}
\author{David Mkrtichian}
\affiliation{National Astronomical Research Institute of Thailand (NARIT), Siripanich Building, 191 Huaykaew Road, Muang District, Chiangmai, Thailand}
\author{Aleksey Shlyapnikov}
\affiliation{Crimean Astrophysical Observatory, 298409, Nauchny, Crimea}

\begin{abstract}
We present and analyze 58 transit light curves of TrES-3b and 98 transit light curves of Qatar-1b observed by Transiting Exoplanet Survey Satellite (TESS), plus two transit light curves of Qatar-1b observed by us using a ground-based 1.23\,m telescope. These light curves are combined with the best-quality light curves taken from the Exoplanet Transit Database (ETD) and literature. The precisely determined mid-transit times from these light curves enable us to obtain the refined orbital ephemerides with improved precision for both hot Jupiters. From the timing analysis, we find an indication for the presence of transit timing variations (TTVs) in both systems. Since the observed TTVs are unlikely to be short-term and periodic, the possibility of additional planets in the orbits close to TrES-3b and Qatar-1b are ruled out. Possible causes of long-term TTVs such as orbital decay, apsidal precession, the Applegate mechanism and line-of-sight acceleration are also examined. However, none of these possibilities are found to explain the observed TTV of TrES-3b. In contrast to this, the line-of-sight acceleration appears to be a plausible explanation for the observed TTV of Qatar-1b. In order to confirm these findings, further high-precision transit and RV observations of both systems would be worthwhile.

\end{abstract}

\keywords{planet-star interactions - stars: individual (TrES-3, Qatar-1) - planets and satellites: individual (TrES-3b, Qatar-1b) - techniques: photometric}



\section{Introduction} \label{sec:intro}

The photometric study of transiting hot Jupiter systems is important for several reasons because it not only helps in improving estimates of their physical and orbital parameters but also provides an opportunity to check for the presence of an additional body in the system through the transit timing variation (TTV) analysis of a known planet \citep[see][]{Maci13,Jiang13,Colln17,Mannaday20,Sariya21,Su21,Athano22}. Moreover, long-term high-precision transit data spanning more than a decade allows tests of the theoretical predictions of orbital decay and apsidal precession in hot Jupiter systems \citep[e.g.,][]{Maci16,Patra17,Patra20,Mannaday20,Athano22}. Orbital decay is predicted for most hot Jupiters because the total angular momentum of the systems are smaller than the critical angular momentum required for a system to achieve a stable state \citep[see,][]{Rasio96,Levrd09,Matsu10}. As the orbital frequency of most hot Jupiters is larger than the host star's rotational frequency, the tidal bulge raised on the host stars by their hot Jupiters exerts torque that leads to orbital decay, by transferring the planet's orbital angular momentum to the stellar rotational angular momentum \citep{Levrd09,Jackson09,Matsu10}. On the other hand, apsidal precession is expected in systems whose hot Jupiters are in at least slightly eccentric ($e>0.003$) orbits \citep{Ragoz09}. The tidal bulge raised on the planet by the host star also exerts torque on the orbit that can lead to apsidal precession \citep[see][]{Ragoz09}. Probing these two phenomena provides an opportunity to infer significant information about the hot Jupiter systems. The orbital decay rate allows the estimation of modified stellar tidal quality factor ($Q^{'}_{\ast}$), a dimensionless parameter that describes the efficiency of energy dissipation in the host star, which is still poorly constrained by observations. The observed apsidal precession rate provides an opportunity to infer the interior density profile of a planet by estimating the planetary tidal Love number ($k_p$) \citep[see][]{Ragoz09,Patra17,Bouma19,Mannaday20}.

In order to address these issues, the Transiting Exoplanet Survey Satellite \citep[TESS;][]{Ricker14}, launched in 2018, is valuable because it is not only discovering new extrasolar planets but also following up the transits of previously known hot Jupiters. Combining ground-based transit data with the high-precision 2-min cadence transit data provided by TESS is extremely helpful for refining the estimates of system parameters \citep[e.g,][]{Cortes20,Ikwut20,Szabo20,South22}, exploring the existence of additional planets \citep[e.g.,][]{Huang18,Teske20,Garai20}, and searching the possibility of long-term trends due to orbital decay, apsidal precession, the Applegate mechanism \citep{Applegate92} and line-of-sight acceleration phenomena in hot Jupiter systems \citep[e.g.,][]{Watson10,Bouma19,Bouma20,South19,Yee20,Battley21,Turner21,Turner22,Wong22}. In the context of orbital decay studies, \citet{Maci16} first reported an orbital decay rate of $-25.60 \pm 4.0$~ms~yr$^{-1}$ for hot Jupiter WASP-12b using transit data spanning a decade. \citet{Patra17} confirmed this but suspected the possibility of apsidal precession. Recent timing analyses performed by \citet{Yee20}, \citet{Turner21} and \citet{Wong22}, including data from TESS and from the literature, provide strong evidence of orbital decay for WASP-12b. The WASP-12 system is currently the only system in which orbital decay has been confirmed directly through transit observations. The second hot Jupiter for which orbital decay has been detected with high significance but could not be confirmed so far is WASP-4b. Using the observed TESS transits and previously published data, \citet{Bouma19} reported an orbital decay rate of $-12.6 \pm 1.2$~ms~yr$^{-1}$ and $\sim 80$~s early arrival of the transits of WASP-4b. \citet{South19} found a similar decay rate for this hot Jupiter using TESS and extensive new ground-based data. Apart from this, they also examined various possible origins of the TTVs such as stellar activity, the Applegate mechanism, apsidal precession and line-of-sight acceleration. Whilst orbital decay and apsidal precession were plausible, the other possibilities were ruled out. Later, a timing and radial velocity (RV) analysis performed by \citet{Bouma20} suggested that the observed decay rate is actually caused by the acceleration of WASP-4b towards the Earth. However, \cite{Turner22} reanalyzed all the TESS, RV and literature data, did not see any indication for the acceleration of WASP-4b towards the Earth, but instead found evidence for the presence of a second planet in the system. In addition to WASP-12b and WASP-4b, there are many hot Jupiters for which a tentative detection of orbital decay has been reported (e.g., WASP-43b: \citealt{Jiang16}; WASP-46b: \citealt{Petrucci18}) but could not be confirmed yet either due to contradictory findings (e.g., WASP-43b: \citealt{Hoyer16,Patra20}; WASP-46: \citealt{Davoudi21}) or due to a lack of observations over sufficient time spans \cite[e.g., KELT-16b, HATS-18b, WASP-18b, WASP-19b, WASP-72b:][]{Patra20,Mancini22}.

In this paper, we have chosen two hot Jupiters, TrES-3b and Qatar-1b, for our timing analysis. The availability of long-term transit data and additional follow-up observations obtained by previous workers are the main reasons for selecting these two systems. TrES-3b was discovered by the Trans-Atlantic Exoplanet Survey \citep{Dono07} orbiting a G-type star ($V=12.4$ mag, $T_{\rm eff} =5720 \pm 150$, $M_\ast = 0.9 \pm 0.15 \ M_{\sun}, R_{\ast}= 0.802 \pm 0.046 \ R_{\sun}$). Qatar-1b was discovered by the Qatar Exoplanet Survey \citep{Alsu11} around a metal-rich K-type star ($V = 13.5$ mag, $T_{\rm eff} =4861 \pm 125$ K, $M{_\ast} = 0.85 \ M_{\sun}, R_{\ast}=0.823 \pm 0.025 \ R_{\sun}$). The transits of both hot Jupiters have been extensively studied in the past to characterize their physical and orbital properties, as well as to examine the presence of additional planets through TTV analysis (e.g., TrES-3b: \citealt{Sozz09,Gibs09,Vanko13,Jiang13,Kund13,Pusk17,Ricc17}; Qatar-1b: \citealt{Covino13,Essen13,Maci15,Misl15,Colln17,Thakur18}). Recently, \cite{Mannaday20} examined the TrES-3 system using 12 new transit light curves with 71 transit light curves from the literature and discussed the possibilities of orbital decay and apsidal precession phenomena for the observed TTV. Because of a statistically insignificant estimation of the orbital decay rate ($\dot{P_q}=-4.1 \pm 3.1$~ms~yr$^{-1}$), they proposed that new data observed by TESS may be useful to confirm their findings. The presence of TTVs has also been reported in the Qatar-1 system by \citet{Su21} using 38 transit light curves. These authors also found a statistically insignificant period decrease of $(-5.9 \pm 5.2) \times {10}^{-10}$~day~epoch$^{-1}$. Because of this, they proposed further follow-up observations of the transits of Qatar-1b to refine their results.

In the context of the discussion above, we note that TESS has observed TrES-3 and Qatar-1 in multiple sectors (sectors 25, 26 and 40 for TrES-3 and sectors 17, 21, 24, 25, 41 and 48 for Qatar-1). We have processed and analyzed these new transit data to further probe the possible TTVs for both hot Jupiters. Apart from the TESS light curves, we have collected more transit light curves from the Exoplanet Transit Database\footnote{http://var2.astro.cz/ETD/} \citep[ETD;][]{Poddany10} and the literature to increase the baseline of the transit observations. Two more transit light curves of Qatar-1b were observed by us using a 1.23\,m telescope. In total, 182 transit light curves of TrES-3b and 228 transit light curves of Qatar-1b, spanning more than decade, are employed in this work.

The remainder of this paper is organized as follows. Section 2 presents the ground-based and TESS observations, as well as transit data taken from the ETD and the literature. Section 3 describes the procedure adopted for analyzing the transit data, and the timing analyses are given in Section 4. The results of this analysis are discussed in Section 5, and concluding remarks are given in Section 6.


\section{Observational Data} \label{sec:observation}

\subsection{Ground-Based Observations}

Two transits of Qatar-1b were observed by us on 2016 August 17 and 2017 March 18 using the 1.23\,m Zeiss telescope at the Calar Alto Observatory (CAHA) in Spain. We used the DLR MKIII 4k$\times$4k CCD camera, which has pixels of size 15\,$\mu m$ and gives a field-of-view of 21.5$\times$21.5 arcmin$^2$. The observations were performed with the telescope defocussed to increase the photometric precision \citep{South09} and with autoguiding. The CCD camera was windowed to decrease the dead time between exposures. Both transits were observed through a Cousins $R$ filter, chosen as it is part of a well-known photometric system and had a high throughput for Qatar-1. The data were reduced in the standard way using the IDL {\sc defot} pipeline \citep{South09,South14}. We obtained bias and flat-field calibrations but did not use them as they had little effect on the final light curve apart from adding a small amount of extra noise \citep[see][]{South14,South19}. We performed aperture photometry with apertures placed manually on a reference image and tracked to follow the stars by cross-correlating each image with the reference image. The resulting light curves were normalized to zero magnitude by fitting a straight line to the out-of-transit data simultaneously with optimising the weights of a set of comparison stars to minimize the scatter in the data taken outside transit. The time stamps were converted from Heliocentric Julian Day (HJD) to Barycentric Julian Day (BJD) on the Barycentric Dynamical Time (TDB) timescale, using IDL procedures from \cite{East10}\footnote{//astroutils.astronomy.ohio-state.edu/time/hjd2bjd.html}. The normalized light curves are shown in Fig.~1 (see Section 3 for details) and the original data points are listed in Table 1.

\begin{figure}[h]
	\includegraphics[width=\columnwidth]{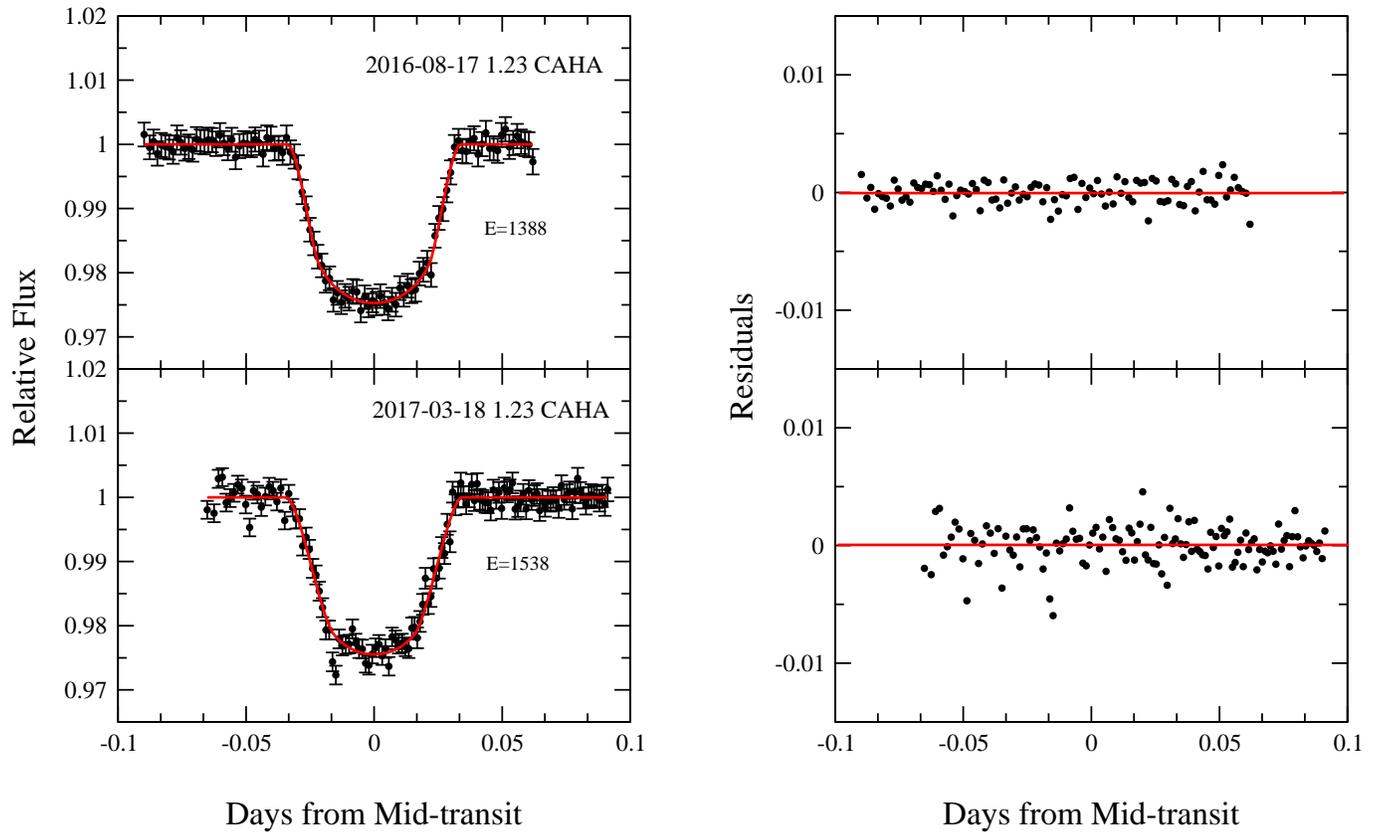}
    \caption{Left panel: The normalized relative flux of Qatar-1 as a function of the time (the offset from mid-transit time and in TDB-based BJD) of two transits observed by us with the 1.23\,m telescope at CAHA. The points are the data and solid lines are the best-fit models. Right panel: The corresponding residuals.}
    \label{fig:1}
\end{figure}

\subsection{TESS Observations}

TESS observed 58 transits of TrES-3b and 98 transits of Qatar-1b up to sector 48. TrES-3 and Qatar-1 are designated as TIC 116264089 and TIC 236887394 in the TESS input catalog\footnote{https://exo.mast.stsci.edu}, respectively \citep[see][]{Stassun19,Paegert21}. The science images taken by TESS were reduced to light curves by the Science Processing Operation Center (SPOC) at NASA Ames \citep{Jenkins16}. For this work the Presearch Data Conditioning (PDC) light curves, optimized to remove instrumental variability through the method discussed by \cite{Smith12} and \cite{Stumpe12,Stumpe14}, were directly retrieved from the Mikulski Archive for Space Telescopes\footnote{https://archive.stsci.edu/} ({\sc MAST}) using the {\sc juliet} package \citep{Espinoza19}. Since {\sc juliet} has a function to discard the data points of non-zero quality flags while extracting the time series data (times, fluxes, and flux errors) from the PDC light curve, only the data associated with zero quality flags were retrieved for our analysis \citep[e.g.,][]{Huber21}. The format of the time of observations, which was initially in TESS Julian Day (TJD$_{\rm TDB}$), was converted to BJD$_{\rm TDB}$ by adding $2,457,000$ \citep{Tenenbaum18}.

To remove the trends that appear in the time series data of TrES-3 and Qatar-1, the out-of-transit (OOT) fluxes were modeled through a Gaussian Process (GP) by masking the in-transit points using the transit ephemeris of TrES-3b and Qatar-1b from \citet{Mannaday20} and \citet{Maci15}, respectively. For this modeling, we used {\sc juliet} package and employed a {\sc celerite} (approximate) Mat\'ern multiplied by exponential kernel \citep{Espinoza19}. The parameters mean out-of-flux (mflux), jitter (in parts-per-million) added in quadrature to the errorbars of instrument (${\sigma}_{\omega}$), amplitude of GP (${\sigma}_{GP}$) and the timescale of the Mat\'ern part of the GP ($\rho_{GP}$) were fitted freely, while the dilution factor for the photometric noise (mdilution) was fixed to unity. The prior distributions for the parameters ${\sigma}_{\omega}$, ${\sigma}_{GP}$ and ${\rho}_{GP}$ were assumed to be uniform, whereas a normal distribution was assumed for mflux \citep[see,][]{Espinoza19}\footnote{https://juliet.readthedocs.io/en/latest/tutorials/gps.html}. The detrended normalized light curves were obtained by dividing the best-fitted GP model flux to the time series data of each sector. As a reference, the trend time series data (top panel), the best-fit GP model flux to OOT data (middle panel) and the detrended times series data (bottom panel) of TrES-3 (for sector 25) and Qatar-1 (for sector 17) are shown in Figs.\ 2 and 3, respectively. In order to get the individual TESS light curve for each transit event of a particular sector, we extracted sections of the detrended normalized light curve within $\pm 0.1$ day around the expected mid-transit times. The extracted light curves of TrES-3 and Qatar-1 are depicted with black points in Figs. A1-5 (see Appendix A). The original data points of these figures are given in Table 1.

\begin{center}
\small\addtolength{\tabcolsep}{4.0pt}
\begin{longtable*}{cccccc}
\caption{The transit light-curve data of TrES-3b and Qatar-1b considered in this work} \label{tab:longtable_captionlable}\\
\hline
\hline
Object Name & Telescope & Epoch & TDB-based BJD & Normalized Flux & Normalized Flux Error\\
\hline
\endfirsthead
TrES-3  & TESS & 3674 & 2458984.70755 & 0.99676 & 0.00380\\
TrES-3  & TESS & 3674 & 2458984.70894 & 0.99744 & 0.00380\\
TrES-3  & TESS & 3674 & 2458984.71033 & 1.00183 & 0.00380\\
\hline
TrES-3  & TESS & 3675 & 2458986.01314 & 1.00439 & 0.00378\\
TrES-3  & TESS & 3675 & 2458986.01591 & 0.99716 & 0.00378\\
TrES-3  & TESS & 3675 & 2458986.01730 & 0.99884 & 0.00377\\
\hline
Qatar-1 & CAHA & 1388 & 2457618.53719 & 1.00153 & 0.00186\\
Qatar-1 & CAHA & 1388 & 2457618.53932 & 0.99953 & 0.00186\\
Qatar-1 & CAHA & 1388 & 2457618.54084 & 1.00043 & 0.00188\\
\hline
Qatar-1 & CAHA & 1538 & 2457831.56558 & 0.99806 & 0.00139\\
Qatar-1 & CAHA & 1538 & 2457831.56825 & 0.99752 & 0.00139\\
Qatar-1 & CAHA & 1538 & 2457831.56994 & 1.00289 & 0.00140\\
\hline
Qatar-1 & TESS & 2196 & 2458765.90522 & 0.99846 & 0.00384\\
Qatar-1 & TESS & 2196 & 2458765.90660 & 0.99770 & 0.00386\\
Qatar-1 & TESS & 2196 & 2458765.90799 & 0.99665 & 0.00385\\
\hline
Qatar-1 & TESS & 2197 & 2458767.32326 & 0.99849 & 0.00385\\
Qatar-1 & TESS & 2197 & 2458767.32465 & 1.00145 & 0.00385\\
Qatar-1 & TESS & 2197 & 2458767.32604 & 0.99830 & 0.00386\\
\hline
\end{longtable*}
Note. This table is available in its entirety in machine-readable form. A portion is shown here for guidance regarding its form and content.
\end{center}

\begin{figure}
\centering
  \begin{tabular}{@{}c@{}}
    \includegraphics[width=\columnwidth]{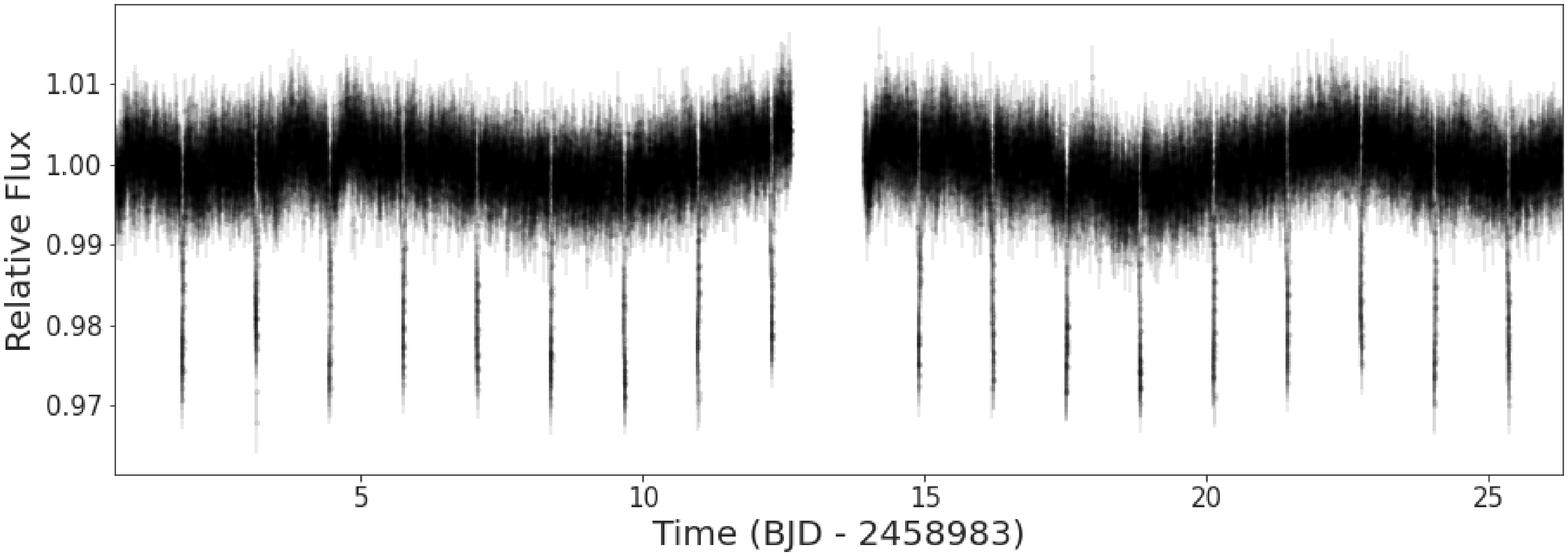}\\
    \includegraphics[width=\columnwidth]{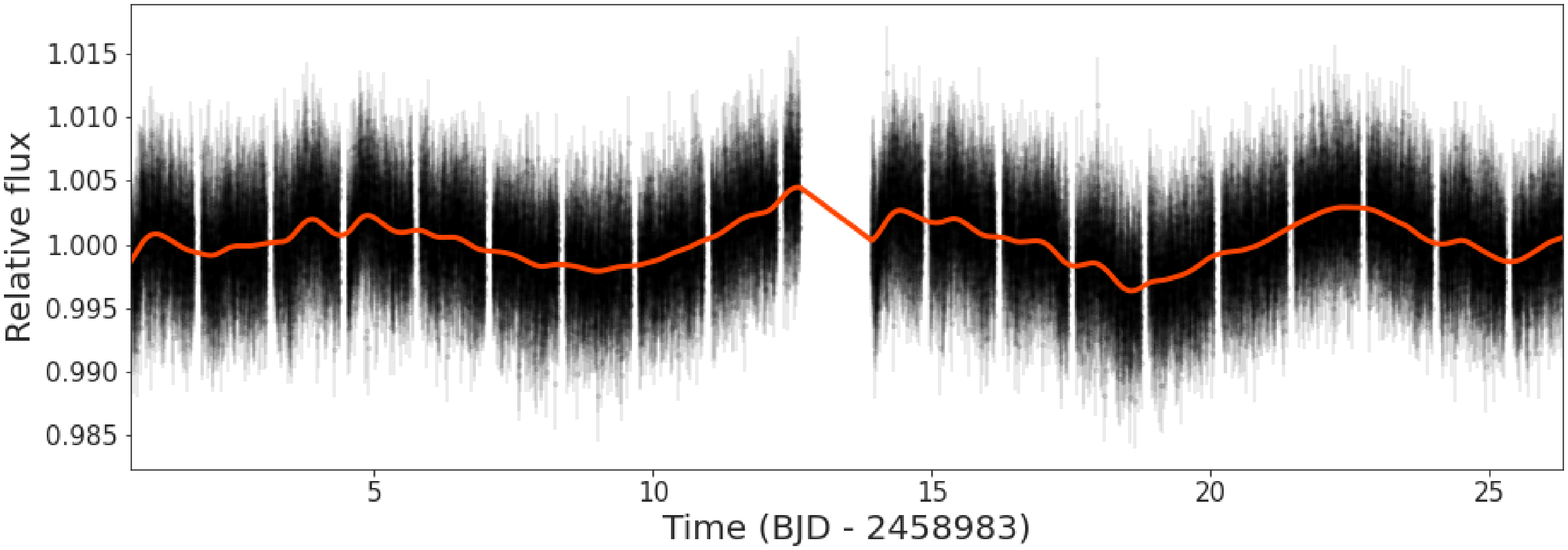}\\
    \includegraphics[width=\columnwidth]{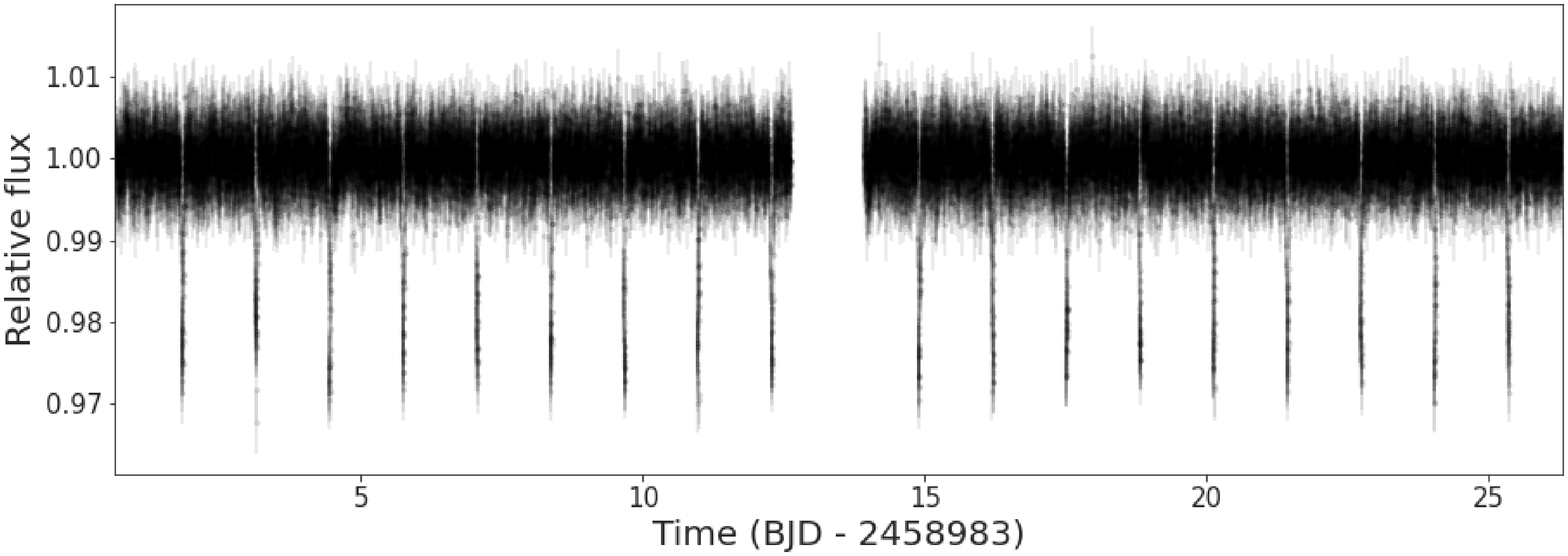} \\
  \end{tabular}
  \caption{Top panel: time series data of TrES-3b observed by TESS within sector 25. Middle panel: corresponding OOT time series data along with the best-fit GP model (red curve). Bottom panel: detrended time series data corresponding to the top panel.}
\end{figure}

\begin{figure}
\centering
  \begin{tabular}{@{}c@{}}
    \includegraphics[width=\columnwidth]{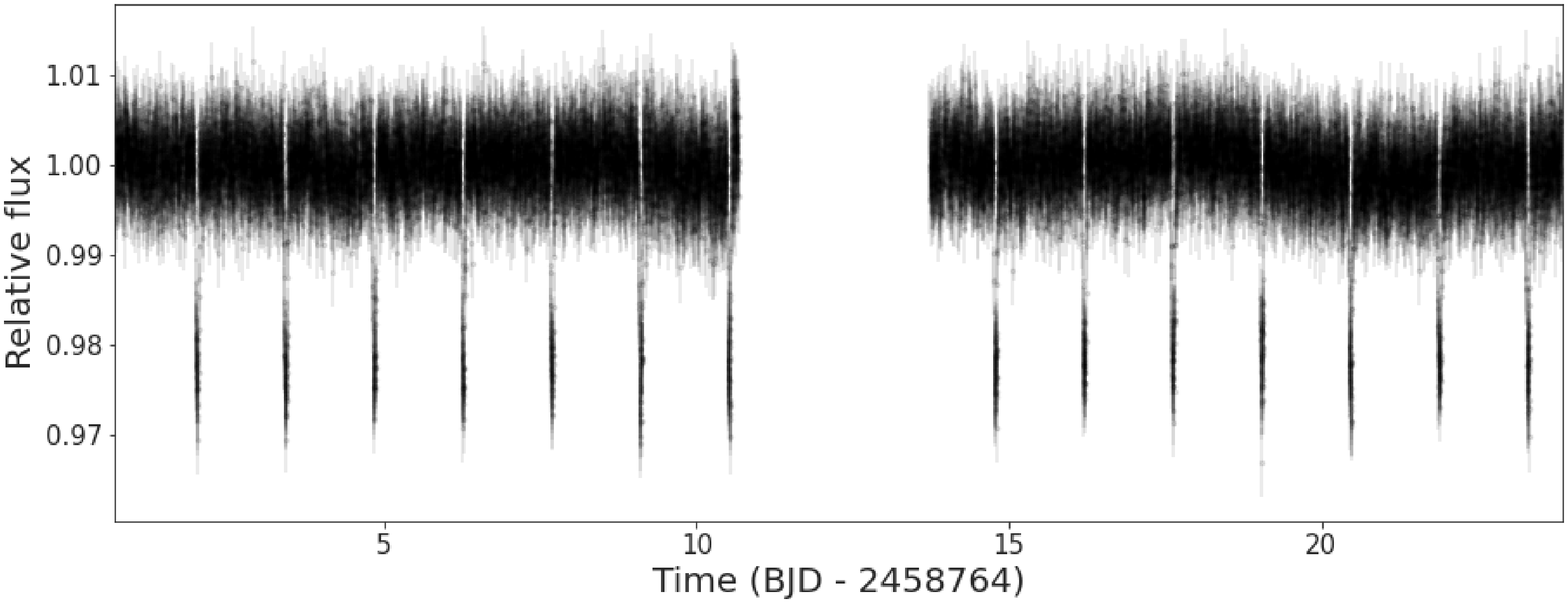}\\
    \includegraphics[width=\columnwidth]{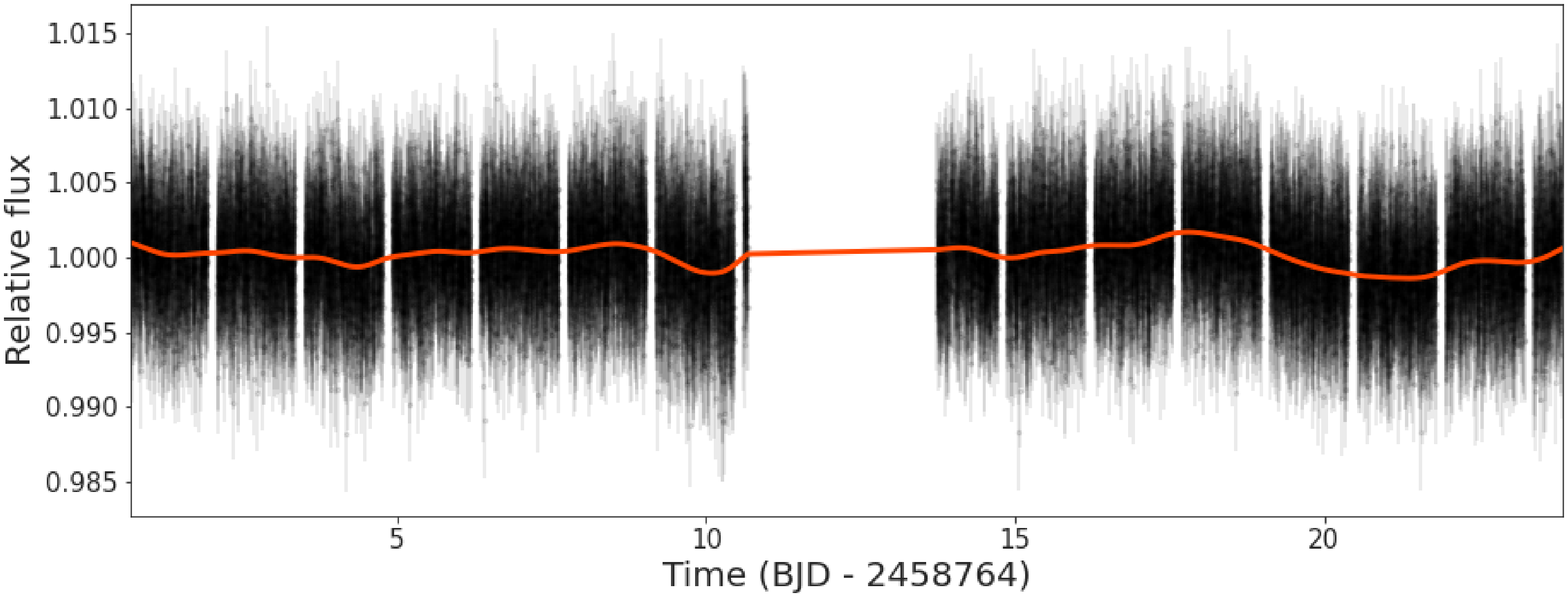}\\
    \includegraphics[width=\columnwidth]{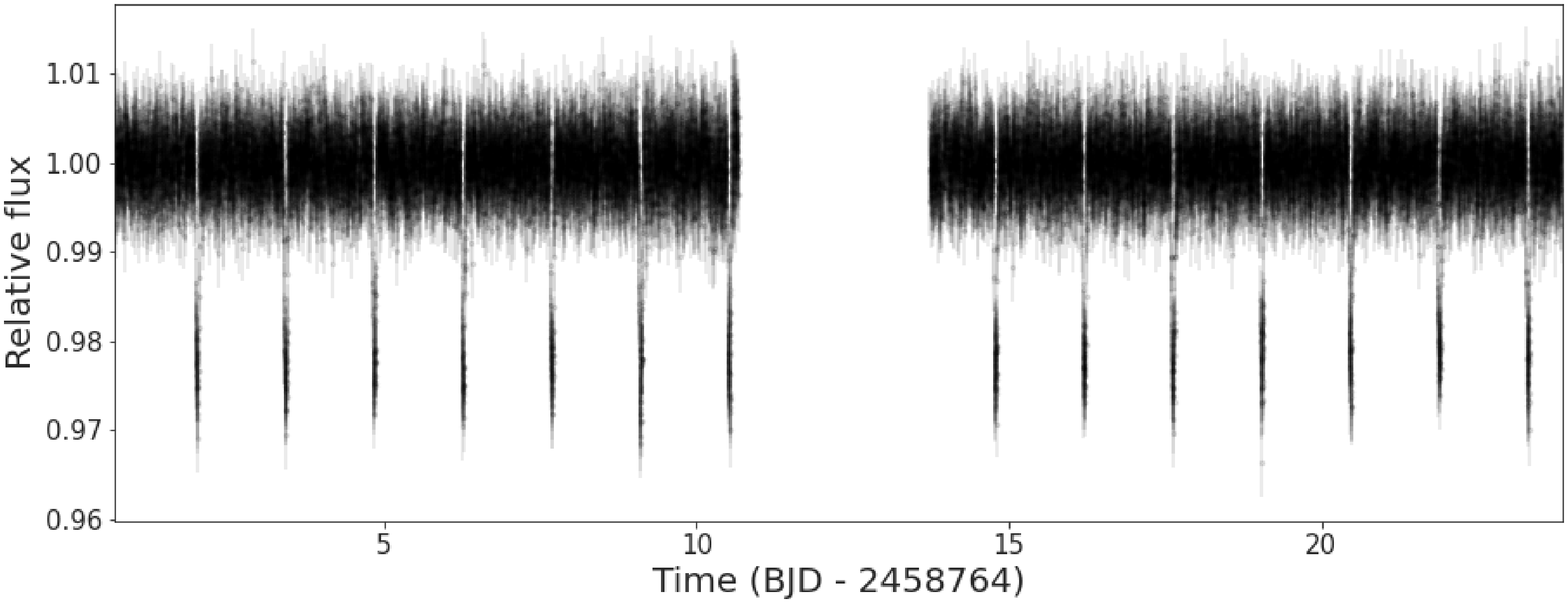} \\

  \end{tabular}
  \caption{Top panel: time series data of Qatar-1b observed by TESS within sector 17. The remaining panels are the same as those mentioned in Fig.~2.}
\end{figure}

\subsection{Observational Data from ETD and the Literature}

In addition to the TESS and CAHA data, 41 transit light curves of TrES-3b and 61 transit light curves of Qatar-1b were taken from the ETD. We included those with quality index $<3$. All the ETD light curves were observed during 2016-2021 by several observers at different observatories across the world. As most of the ETD light curves were not normalized, we normalized them by fitting a linear function of time to the OOT parts and also converted their time stamps from JD or HJD to BJD$_{\rm TDB}$ using the same tool as discussed above. Apart from these light curves, published light curves of both targets available in the literature or available with us were also considered. The details of all the transit light curves of both hot Jupiters employed in this paper are given in Table 2.

\begin{table}
\begin{center}
\caption{Details of the transit light curves of TrES-3b and Qatar-1b considered in this work.}
\label{tab:1}
\begin{tabular}{lllc}
\hline
Object & Number of light  & Sources & Total number of \\
name &  curves taken & & light curves \\
\hline
TrES-3b & 58 & TESS&\\
& 41 & ETD \citep{Poddany10} & 182\\
& 83 & \citet{Mannaday20}&\\
\hline
Qatar-1b & 98 & TESS& \\
& ~2 & Our observation&\\
& 61 & ETD \citep{Poddany10}&\\
& 11 & \citet{Misl15} & 228 \\
& 18 & \citet{Colln17}&\\
& 38 & \citet{Su21}&\\
\hline
\end{tabular}
\end{center}
\end{table}


\section{Light Curve Analysis}

For analyzing the transit light curves of TrES-3b and Qatar-1b, we used the Transit Analysis Package \citep[TAP;][]{Gaz12}. All the light curves of each system were loaded into TAP, separately, to determine the transit parameters such as ratio of planet to star radius (${R_p /R_\ast}$), mid-transit time (${T_m}$), orbital inclination ({\it i}), scaled semi-major axis ($a/R_\ast$), linear and quadratic limb-darkening (LD) coefficients ($u_1$, $u_2$).  For each light curve analysis, we used five MCMC chains with a lengths of $10^5$ links each and adopted exactly the same parameter fitting procedure as described in \citet{Mannaday20} for TrES-3b and in \citet{Su21} for Qatar-1b. To start the TAP run, the initial values of the parameters ${R_p /R_\ast}$, {\it i}, and $a/R_\ast$ were taken from \citet{Sozz09} for TrES-3 and \citet{Maci15} for Qatar-1.

For the TESS light curves of TrES-3b and Qatar-1b, the values of the linear and quadratic LD coefficients ($u_{1}$, $u_{2}$) were taken from the tables of \citet{Clar17}. In the case of ETD light curves of TrES-3b observed in clear, {\it V}, {\it I}, and {\it R} filters, the values of $u_{1}$ and $u_{2}$ were directly adopted from Table 4 of \citet{Mannaday20}, who derived these coefficients using the \textsc{jktld}\footnote{\textsc{jktld} is available from http://www.astro.keele.ac.uk/jkt/codes.html.} code \citep{South15}. For the light curves of Qatar-1b observed in the {\it V}, {\it R} and {\it I} filters, we followed \citet{Su21} and linearly interpolated the values of $u_{1}$ and $u_{2}$ from the tables of \citet{Clar11} using the {\sc exofast}\footnote{https://astroutils.astronomy.osu.edu/exofast/limbdark.shtml} package \citep{East13} with the stellar parameters effective temperature ($T_{\rm eff} = 4910$~K), surface gravity ($log_{g} = 4.55$), and metallicity ([Fe/H] $= 0.2$) \citep{Covino13,Maci15}. Since a clear filter covers the {\it V} and {\it R} bands, the LD coefficients for the light curves of Qatar-1b observed through a clear filter were taken as the average of their values in the {\it V} and {\it R} filters \citep[see][]{Maci13,Mannaday20}. For the light curves of Qatar-1b observed using the Gunn $g$, $r$ and $z$ filters by \citet{Misl15}, the values of the LD coefficients calculated for the SDSS {\it g}, {\it r} and {\it z} filters were used, respectively. We followed \citet{Colln17} and used the values of LD coefficients calculated with the \textit{Kepler} and $R$ filters for their Clear with Blue Block (CBB; a high pass filter with cutoff at $\sim 500$~nm) and an open filter, respectively. To maintain the homogeneity in analyzing and deriving the transit parameters as those obtained in \citet{Mannaday20} and \citet{Su21}, the different packages were used to calculate the LD coefficients for TrES-3 and Qatar-1 stars. The initial values of the LD coefficients used for TrES-3 and Qatar-1 for different filters are listed in Table 3.

After the successful completion of an MCMC run of TAP, the initial 10\% of the drawn samples of model parameters were discarded as the burn-in stage and the remaining samples of the model parameters were used for Bayesian parameter extraction. The 50-th percentile level (median) of the posterior probability distribution for each model parameter was interpreted as the best-fit value. Moreover, the 15.9 and 84.1 percentile levels (i.e. 68\% credible intervals) of the posterior probability distribution were taken to be the lower and upper $1\sigma$ uncertainties, respectively. The values of the model parameters ${R_p /R_\ast}$, ${T_m}$, {\it i}, $a/R_\ast$, $u_1$, $u_2$ and their  $1\sigma$ uncertainties derived from the TESS light curves of TrES-3b are given in Table 4, whereas those derived from our and the TESS light curves of Qatar-1b are given in Table 5. The red curves overplotted in the left panels of Fig.~1 and Figs.~A1-5 are the best-fit transit models. To perform a homogeneous and precise timing analysis \citep[e.g.,][]{Petrucci18,Mannaday20}, 83 mid-transit times of TrES-3b and 38 mid-transit times of Qatar-1b were directly taken from \citet{Mannaday20} and \citet{Su21}, respectively. The mid-transit times of TrES-3b and Qatar-1b derived in this paper along with those taken from \citet{Mannaday20} and \citet{Su21} are gathered in Table 6 and Table 7, respectively.

\begin{table*}
\begin{center}
\caption {The theoretical limb-darkening coefficients}
\label{tab:1}
\begin{tabular}{clcc}
\hline
Object Name & Filter & $u_1$ & $u_2$\\
\hline
\multirow{5}{*}{TrES-3} & {\it V} $^{a}$ & 0.4378 & 0.2933\\
& {\it R} $^{a}$ & 0.3404 & 0.3190\\
& {\it I} $^{a}$ & 0.2576 & 0.3186\\
& clear $^{a}$ & 0.3891 & 0.3062\\
& TESS$^{b}$ & 0.3799 & 0.2051\\
\hline
\multirow{9}{*}{Qatar-1} & {\it V}$^{c}$ & 0.6971 & 0.0880\\
& {\it R} $^{c}$ & 0.5579 & 0.1600\\
& {\it I}$^{c}$ & 0.4304 & 0.1962\\
& Sloan \ {\it g}$^{c}$ & 0.8403 & -0.0061\\
& Sloan \ {\it r}$^{c}$ & 0.5949 & 0.1479\\
& Sloan \ {\it z}$^{c}$ & 0.3715 & 0.2110\\
& Clear$^{d}$ & 0.6275 & 0.1240\\
& CBB (Kepler filter)$^{c}$ & 0.6025 & 0.1291\\
& TESS$^{b}$ & 0.4518 & 0.1870\\
\hline
\end{tabular}\\
\end{center}
{Notes:
$^a$ $u_1$ and $u_2$ directly adopted from \citet{Mannaday20}.\\
$^b$ $u_1$ and $u_2$ taken from the tables of \citet{Clar17}.}\\
$^c$ Calculated using {\sc exofast} with ${T}_{\rm eff}=4910$~K, $\log{g}=4.55$ and [Fe/H] $=0.2$.\\
$^d$ Calculated as the average of their value in the {\it V} and {\it R} filters.\\
\end{table*}

\begin{center}
\small\addtolength{\tabcolsep}{2.0pt}
\begin{longtable*}{cp{3.5cm}cp{1.8cm}cp{1.8cm}cp{1.5cm}cp{1.5cm}cp{1cm}c}
\caption{The  Best-fit Values of Parameters ${T}_{m}$, $\it {i}$, ${a/{R}_{\ast}}$,${{R}_{p}/{R}_{\ast}}$,  ${{u}_{1}}$, and ${{u}_{2}}$ for 58 TESS Transit Light Curves of TrES-3b \label{tab:longtable_captionlable}}\\
\hline
\hline \multicolumn{1}{c}{Epoch} & \multicolumn{1}{c}{${T}_{m}$} & \multicolumn{1}{c}{$\it i$} & \multicolumn{1}{l}{$a/{R}_{\ast}$} & \multicolumn{1}{c}{R$_p$/R$_\ast$} & \multicolumn{1}{c}{u$_1$} & \multicolumn{1}{c}{u$_2$}\\
(E) & \ \ \ \ \ (BJD$_{\rm TDB}$) & (deg) &  &  &  & \\
\hline
\endfirsthead
		3674 & $2458984.83927^{+0.00070}_{-0.00068}$ & $81.84^{+0.14}_{-0.14}$ & $5.921^{+0.052}_{-0.051}$ & $0.1645^{+0.0072}_{-0.0063}$ & $0.365^{+0.050}_{-0.050}$ & $0.232^{+0.050}_{-0.050}$\\
		3675 & $2458986.14594^{+0.00080}_{-0.00077}$ & $81.86^{+0.15}_{-0.14}$ & $5.913^{+0.052}_{-0.052}$ & $0.1598^{+0.0074}_{-0.0067}$ & $0.363^{+0.050}_{-0.050}$ & $0.229^{+0.050}_{-0.050}$\\
\hline
\end{longtable*}
Note. This table is available in its entirety in machine-readable form. A portion is shown here for guidance regarding its form and content.
\end{center}

\begin{center}
\small\addtolength{\tabcolsep}{2pt}
\begin{longtable*}{cp{3.5cm}cp{1.8cm}cp{1.8cm}cp{1.5cm}cp{1.5cm}cp{1cm}c}
\caption{The  Best-fit Values of Parameters ${T}_{m}$, $\it {i}$, ${a/{R}_{\ast}}$,${{R}_{p}/{R}_{\ast}}$,  ${{u}_{1}}$, and ${{u}_{2}}$ for our two and 98 TESS Transit Light Curves of Qatar-1b \label{tab:longtable_captionlable}}\\
\hline
\hline \multicolumn{1}{c}{Epoch} & \multicolumn{1}{c}{${T}_{m}$} & \multicolumn{1}{c}{$\it i$} & \multicolumn{1}{l}{$a/{R}_{\ast}$} & \multicolumn{1}{c}{R$_p$/R$_\ast$} & \multicolumn{1}{c}{u$_1$} & \multicolumn{1}{c}{u$_2$}\\
(E) & \ \ \ \ \ (BJD$_{\rm TDB}$) & (deg) &  &  &  & \\
\hline
\endfirsthead
		1388 & $2457618.62696^{+0.00024}_{-0.00023}$ & $85.85^{+0.92}_{-0.71}$ & $6.89^{+0.37}_{-0.30}$ & $0.1464^{+0.0028}_{-0.0033}$ &
$0.565^{+0.046}_{-0.049}$ & $0.165^{+0.049}_{-0.049}$\\
		1538 & $2457831.63075^{+0.00033}_{-0.00029}$ & $83.04^{+0.46}_{-0.41}$ & $6.01^{+0.20}_{-0.18}$ & $0.1555^{+0.0024}_{-0.0024}$ &
$0.539^{+0.049}_{-0.049}$ & $0.146^{+0.050}_{-0.049}$\\
\hline
\end{longtable*}
Note. This table is available in its entirety in machine-readable form. A portion is shown here for guidance regarding its form and content.
\end{center}

\begin{center}
\small\addtolength{\tabcolsep}{-3pt}
\begin{longtable*}{cp{3.8cm}cp{4.0cm}cp{3.5cm}lll}
\caption{Mid-transit Times $({T}_{m})$ and Timing Residuals (O-C) for 182 Transit Light Curves of TrES-3b} \label{tab:long}\\
\hline
\hline \multicolumn{1}{c}{Epoch} & \multicolumn{1}{c}{${T}_{m}$} & \multicolumn{1}{c}{O-C} & \multicolumn{1}{l}{Transit Source} & \multicolumn{1}{l}{Timing Source} & Date Excluded\\
(E)	&	\ \ \ \ \ \ (BJD$_{\rm TDB}$)	& (days)	& & & from timing analysis\\
\hline
\endfirsthead
 0 & $2454185.91110^{+0.00021}_{-0.00021}$ 		& -0.0000490 	& \citet{Sozz09} & \citet{Mannaday20} & \\
10 & $2454198.97359^{+0.00057}_{-0.00066}$ 		& 0.0005782 	& \citet{Sozz09} & \citet{Mannaday20} &\\
\hline
\end{longtable*}
Note. This table is available in its entirety in machine-readable form.\\
References. \citet{Sozz09,Gibs09,Coln10,Lee11,Jiang13,Vanko13,Turnr13,Kund13,Pusk17,Ricc17,Mannaday20}
\end{center}

\begin{center}
\small\addtolength{\tabcolsep}{-3pt}
\begin{longtable*}{cp{3.8cm}cp{4.0cm}cp{3.5cm}lll}
\caption{Mid-transit Times $({T}_{m})$ and Timing Residuals (O-C) for 228 Transit Light Curves of Qatar-1b} \label{tab:long}\\
\hline
\hline \multicolumn{1}{c}{Epoch} & \multicolumn{1}{c}{${T}_{m}$} & \multicolumn{1}{c}{O-C} & \multicolumn{1}{l}{Transit Source} & \multicolumn{1}{l}{Timing Source} & Date Excluded\\
(E)	&	\ \ \ \ \ \ (BJD$_{\rm TDB}$)	& (days)	& & & from timing analysis\\
\hline
\endfirsthead
0  & ${2455647.63228}^{+0.00031}_{-0.00033}$   & -0.0008948 & \citet{Essen13} & \citet{Su21} & \\
45 & ${2455711.53484}^{+0.00019}_{-0.00021}$   & 0.0005704 	& \citet{Covino13} & \citet{Su21} &\\
\hline
\end{longtable*}
Note. This table is available in its entirety in machine-readable form.\\
References. \citet{Essen13,Maci15,Misl15,Colln17,Su21}.
\end{center}


\section{Transit Timing Analysis}

\subsection{New Ephemeris}

Although 182 mid-transits times of TrES-3b and 228 mid-transits times of Qatar-1b are listed in Tables 6-7, we have selected mid-transit times with uncertainties less than 1 minute for precise timing analysis \citep[e.g.,][]{Maci13,Petrucci18,Shan21}. A total of 129 mid-transit times of TrES-3b and 184 mid-transit times of Qatar-1b passed this selection criterion.

We derived new linear ephemerides for both hot Jupiters by fitting a linear model,
\begin{equation}
T^{c}_{m} (E) = T_0 + EP,
\end{equation}
to their precise mid-transit times as a function of epoch $E$ using the $emcee$ MCMC sampler implementation \citep{Foreman13}, where ${T}^{c}_{m}$, $E$, {\it P}, and ${T}_{0}$ are the calculated mid-transit time, epoch, orbital period, and mid-transit time of reference epoch ($E = 0$)\footnote{The first transit of TrES-3b and Qatar-1b observed by \citet{Sozz09} and \citet{Essen13} were considered as $E=0$, respectively}, respectively. To run the MCMC chain for model fitting, we followed \cite{Mannaday20} and assumed a Gaussian likelihood, as well as imposing uniform priors on the model parameters, $P$ and $T_0$ (see Table 8). We used 100 walkers and ran 300 steps per walker as initial burn-in to set the step size of each parameter. Followed by this, the code was further run for a final $20,000$ steps per walker to sample the posterior probability distributions of the model parameters. In order to assess the convergence and sampling of the MCMC chains, we estimated the mean acceptance fraction ($a_f$), the integrated autocorrelation time ($\tau$) and the effective number of independent samples ($N_{\rm eff}$), which are listed in Table 8. These parameters indicate efficient convergence and good sampling of the MCMC chains, as the estimated value of $a_f$ lies within the ideal range of 0.2--0.5 and the value of $N_{\rm eff}$ is found to be larger than its minimum threshold value of 50 per walker set in our MCMC analysis \citep[see,][]{Goodman10,Foreman13,Cloutier16,Stefansson17,Mannaday20}.

Before performing the Bayesian parameter extraction from the drawn samples of the posterior probability distributions, we discarded $\sim 2\tau$ steps from the $20,000$ steps of each walker as a final burn-in to get rid of the strongly correlated parameters. The median and 68\% credible intervals of the remaining samples of the posterior probability distribution of each model parameter are considered as the best-fit value and its lower and upper $1\sigma$ uncertainties. The best-fit parameters and their $1\sigma$ uncertainties derived from linear model fit are listed in Table 8 for both targets. Moreover, the values of ${\chi}^{2}$, ${\chi}^{2}_{red}$ and the Bayesian Information Criterion ($BIC= {\chi}^{2} + k\log{N}$, where $k$ is the number of free parameters and $N$ is the number of data points) obtained corresponding to the best-fit model parameters are also listed in this table. The derived values of linear ephemeris for both the hot Jupiters are consistent with results available in the literature. However, they are estimated with improved precision. Comparing the mid-transit times of TrES-3b derived from the TESS light curves with those estimated using the linear ephemeris of \citet{Mannaday20}, we have found a median difference of $\sim 32.26$~s between these two timings. From this it appears that the TESS transits of TrES-3b occur later than the predictions of linear ephemeris of \citet{Mannaday20}. Similarly, the mid-transit times of Qatar-1b derived from TESS light curves also appear to occur $\sim 113.65$~s later than predicted by the linear ephemeris of \citet{Su21}. Because of these timing differences and the above poor model fittings to timing data with ${\chi}^{2}_{red}>1$, we suspect the possibility of TTVs in both hot Jupiters. To explore this further, we obtained the timing residuals, $O-C$, by subtracting the mid-transit times calculated using the derived ephemeris, $T^{c}_{m}$, from the observed mid-transit times, $T_m$, for all the considered epochs, E. The estimated timing residuals are given in Table 6 (for TrES-3b) and Table 7 (for Qatar-1b), and are also depicted as a function of epoch $E$ in Fig.\ 4. The $rms$ of the timing residuals of TrES-3b is $\sim 65.56$~s, while that of Qatar-1b is $\sim 46.44$~s. \\

\subsection{A Periodicity Search for Additional Planets}

To probe the periodicity in the timing residuals of both hot Jupiters that may be induced due to presence of an additional body, we computed the generalized Lomb-Scargle periodograms \citep[GLS;][]{Zechme09} for their timing residuals in the frequency domain. The resulting periodogram of TrES-3b is shown in the top panel of Fig.~5. In this periodogram, the highest peak of power 0.1285 was found at the frequency of $0.01297$ cycle/epoch and the false alarm probability (hereafter FAP) corresponding to this highest power was $10\%$. This FAP was determined empirically by randomly permuting the timing residuals to the observing epochs using a bootstrap resampling method with ${10}^5$ trials. The periodogram computed for the timing residuals of Qatar-1b is depicted in the bottom panel of Fig.~5. Here, the highest peak power of 0.1640 was found at the frequency of $0.155942$ cycle/epoch and the corresponding FAP was $24\%$. In both periodograms, the estimated FAPs are found to be far below the threshold level of FAP=5\%, which means that short-term periodic TTVs in the timing residuals of both the hot Jupiters are not detected. Therefore, the possibility of an additional body in the orbits close to both the hot Jupiters is ruled out. These findings are fully consistent with the previous results available in the literature (e.g., TrES-3: \citealt{Vanko13,Kund13,Pusk17,Mannaday20}, Qatar-1b: \citealt{Maci15,Thakur18,Colln17,Su21}). The absence of short-term periodic TTVs in the timing residuals of both the hot Jupiters motivate us to explore the possibility of long-term TTVs, which may be induced due to either orbital decay or apsidal precession.

\begin{figure*}
\centering
  \begin{tabular}{@{}c@{}}
    \includegraphics[width=0.9\textwidth]{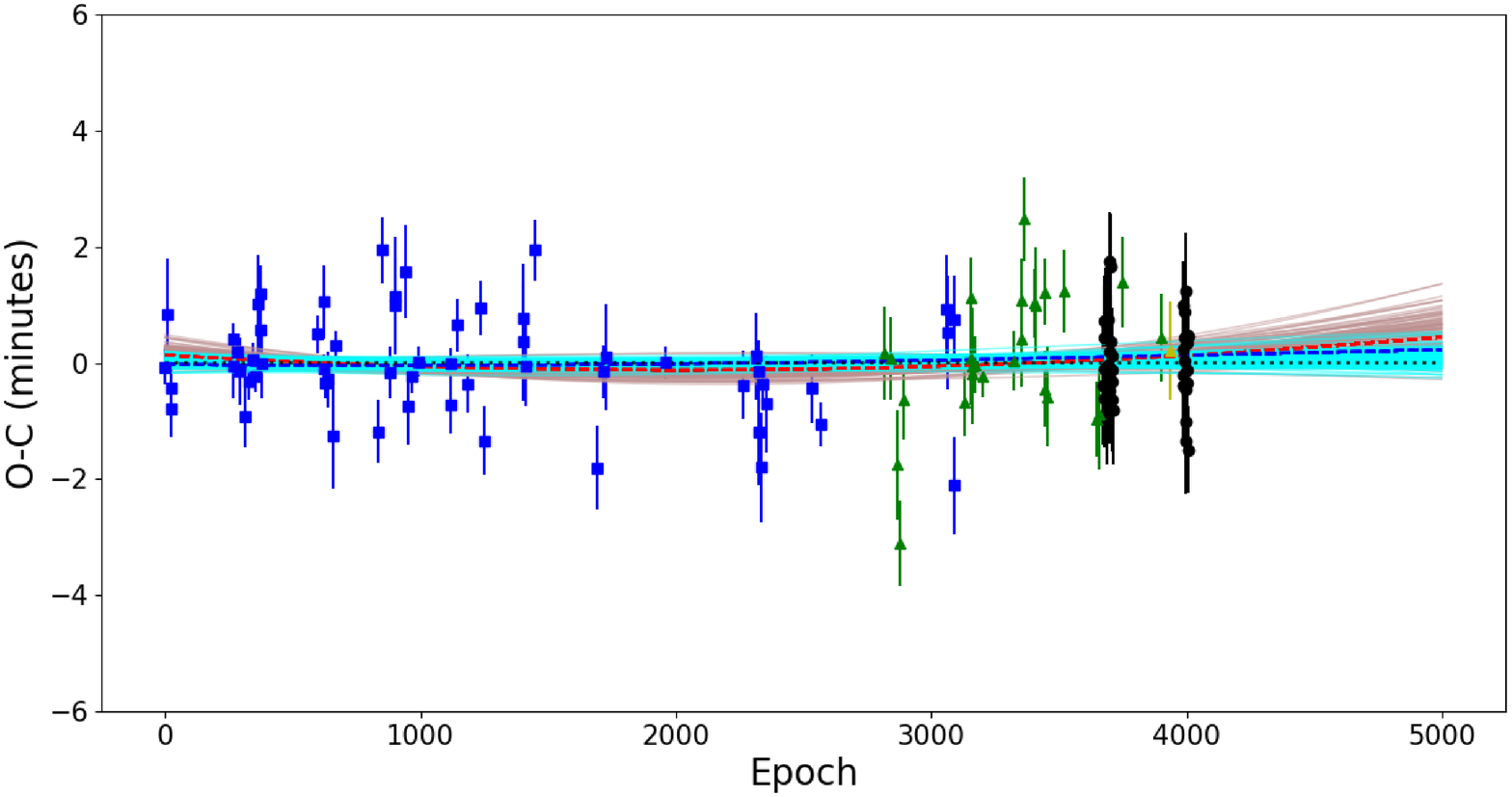}\\
    \includegraphics[width=0.9\textwidth]{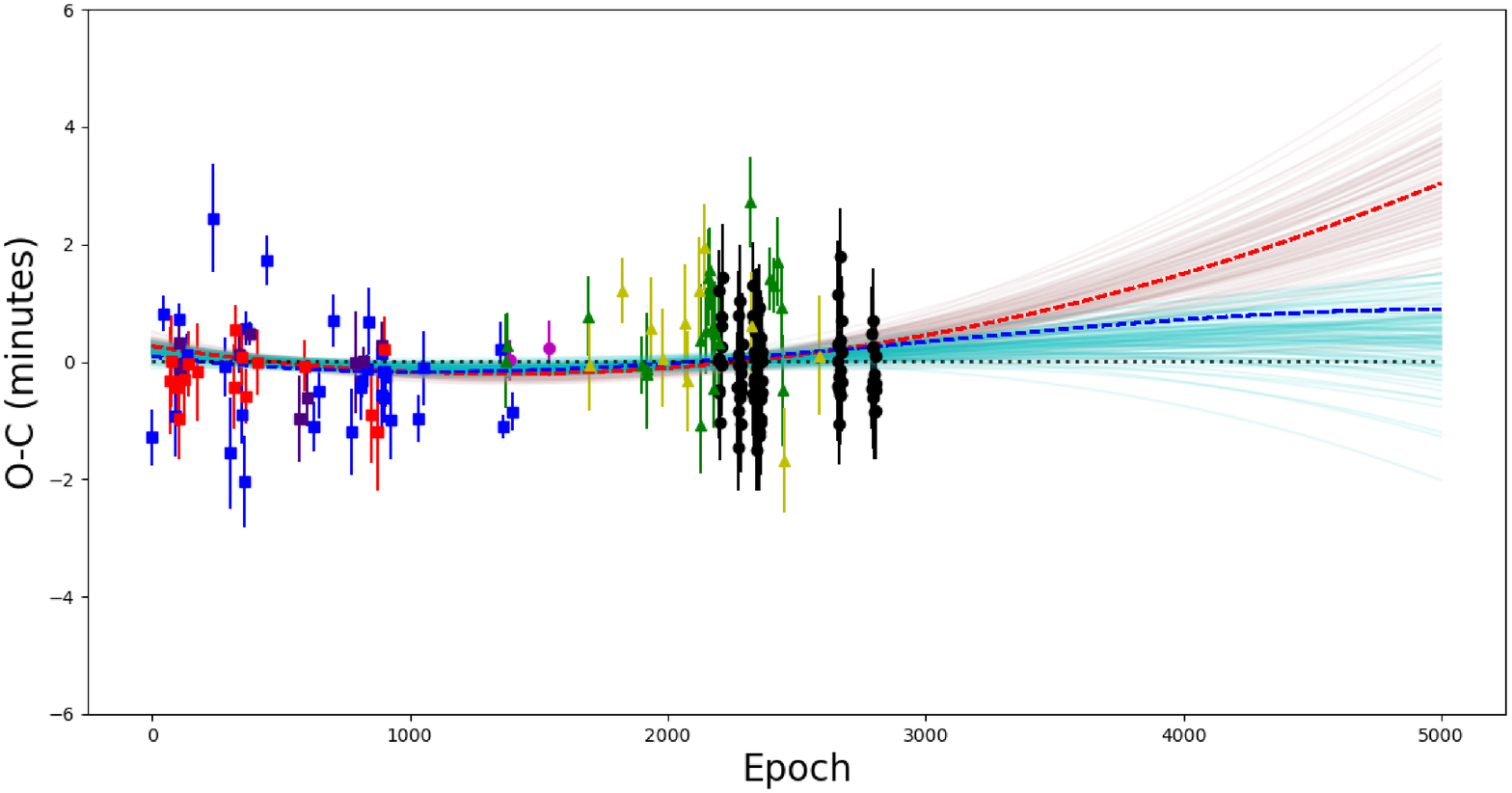}
\end{tabular}
  \caption{Top panel: $O-C$ diagram for the analysis of 129 mid-transit times of TrES-3b. Black filled circles are the data from TESS observations, blue filled squares are from \citet{Mannaday20}, green and yellow filled triangles are from the quality 1 and quality 2 data of ETD. The dotted black line, dashed red and blue curves indicate the linear, orbital decay and apsidal precession models, respectively. The lines are 100 random draws from the posteriors of the orbital decay (brown) and apsidal precession (cyan) models. The scenarios from both the models remain statistically indistinguishable in the time covered by current observations. The models are extrapolated for next $\sim 4.5$ years to illustrate the broad spectrum of possible solutions. Bottom panel: $O-C$ diagram for the analysis of 184 mid-transit times of Qatar-1b. Magneta filled circles are for the data from our observations, black filled circles are for TESS observations, blue filled squares are from \citet{Su21}, indigo filled squares are from \citet{Misl15}, and red squares are from \citet{Colln17}. Other features of this panel are the same as described for the top panel.}
\end{figure*}

\begin{figure}
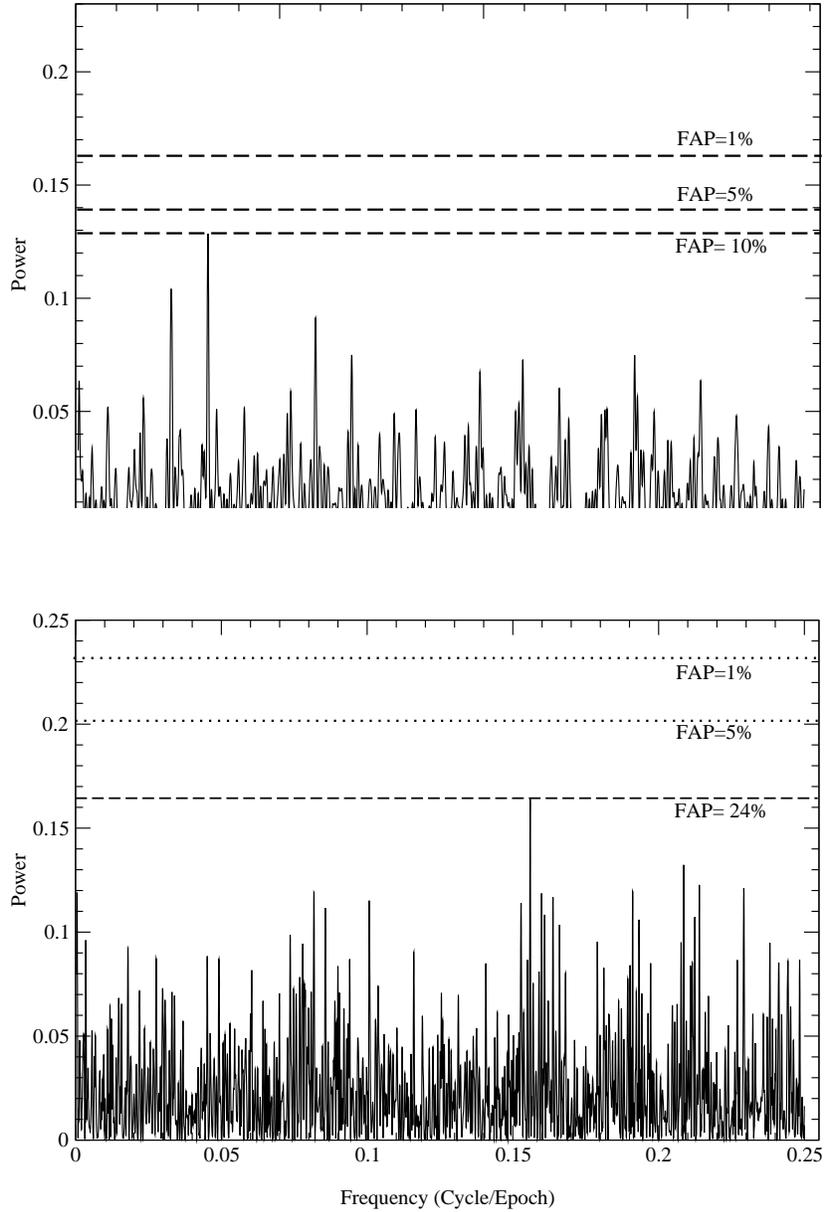

\centering
 \begin{tabular}{@{}c@{}}
   \includegraphics[width=.6\textwidth]{fig5_a.eps}\\
   \includegraphics[width=.6\textwidth]{fig5_b.eps}
 \end{tabular}
  \caption{Generalized Lomb-Scargle periodograms computed for the timing residuals of TrES-3b (top panel) and Qatar-1b (bottom panel). The dashed line in each periodogram indicates the FAP level of the highest power peak. The dotted lines from top to bottom indicate the threshold levels of FAP=1\% and FAP=5\%, respectively.}
\end{figure}

\subsection{A Search for Orbital Decay}

As mentioned above, the long-term TTVs in the hot Jupiter systems may be produced by orbital decay \citep{Levrd09,Matsu10,Penev18}. We thus explored this possibility for the TrES-3 and Qatar-1 systems. To do so, we followed \citet{Maci21} and fitted the following orbital decay model to the timing data of both the hot Jupiters:
\begin{equation}
T^{c}_{q} (E) = {T}_{q0} + P_q  E + \ \frac{1}{2} \ \frac{dP_q}{dE} \ {E}^2,
\end{equation}
where $E$ is the epoch number, ${T}_{q0}$ is the mid-transit time at $E=0$, $P_q$ is the orbital period, $\frac{dP_q}{dE}$ is the change of orbital period in each orbit, and $T^{c}_{q}(E)$ is the calculated mid-transit time. Here, we adopted the same procedure as employed in the case of linear model fit (see Section 3.1), except for running $32,000$ steps per walker in the MCMC for deriving the best-fit ephemeris for the orbital decay model (i.e., $P_q$, $T_{q0}$, $\frac{dP_q}{dE}$). The best-fitting orbital decay ephemerides are given in Table 8. The timing residuals of the orbital decay ephemeris model was obtained by subtracting the mid-transit times calculated using the linear ephemeris, $T^{c}_{m} (E)$, from those estimated using the orbital decay ephemeris, $T^{c}_{q} (E)$. The resulting timing residuals, $T^{c}_{q} (E)-T^{c}_{m}(E)$, are plotted as a function of epoch with red dashed curves in the $O-C$ diagram of each hot Jupiter (see Fig.4). The brown solid lines depicted in the $O-C$ diagrams are 100 random draws from the posteriors of the orbital decay model, which are extrapolated for the next $\sim 4.5$ years to illustrate the future trend of the decay scenario. Using the derived values of $P_q$ and $\frac{dP_q}{dE}$ for TrES-3b and Qatar-1b, the period derivatives are calculated using $\dot {P_q} =\frac{1}{P_q}\frac{dP_q}{dE}$ and are found to be $\sim 2.12 \pm 1.92$~ms~yr$^{-1}$ and $\sim 7.60 \pm 3.19$~ms~yr$^{-1}$, respectively. Since these positive values of $\dot {P_q}$ cannot be attributed to the orbital decay phenomenon, the observed change in the orbital period may be caused by another phenomenon such as a third body in a wider orbit or apsidal precession \citep{Patra20}.

\subsection{A Search for Apsidal Precession}

The previous theoretical work of \citet{Ragoz09} predicts that several hot Jupiters such as WASP-4b, WASP-12b, CoRoT-1b, OGLE-TR-56b and TrES-3b are good candidates to explore the possibility of apsidal precession if their orbits are at least slightly eccentric. Since the eccentricity of Qatar-1b is reported to be greater than the minimum threshold value of 0.003 ($e=0.021^{+0.011}_{-0.010}$: \citealt{Covino13}; $e \sim 0.012$: \citealt{Bonomo17}), this hot Jupiter is a good candidate for apsidal precession study.  As the apsidal precession phenomenon can also produce long-term periodic TTVs in hot-Jupiter systems \citep[see,][]{Ragoz09,Maci16,Patra17,Bouma19,Yee20,Athano22}, we probed the possibility of this phenomenon in the TrES-3 and Qatar-1 systems by adopting the following apsidal precession model \citep[Equation 3 of][which was derived from Equations 7, 9 and 10 of \citealt{Patra20}]{Mannaday20}:
\begin{equation}
T_{ap} (E) = T_{ap0} + P_{s} E -  \frac{e {P_s} \cos{({\omega}_0 + E \frac{d\omega}{dE})}}{\pi(1-\frac{\frac{d\omega}{dE}}{2 \pi})},
\end{equation}
where $E$ is the epoch, $T_{ap}(E)$ is the calculated mid-transit time, $P_{s}$ is the sidereal period, $e$ is the orbital eccentricity, $\omega$ is the argument of periastron, $\omega_0$  is the argument of periastron at epoch zero ($E=0$), and $\frac{d\omega}{dE}$ is the precession rate of periastron. We fitted the above model to the mid-transit times of both the hot Jupiters as a function of epoch by following the procedure as described in Section 3.1, except for running $200,000$ steps per walker in the MCMC chains to determine the best-fit ephemeris of the apsidal precession model (i.e, $P_s$, $T_{a0}$, $e$, $\omega_0$, $\frac{d\omega}{dE}$). Initially, we tried to fit the model by considering wide ranges of uniform priors for the model parameters $e$, $\omega_0$ and $\frac{d\omega}{dE}$, but could not find reliable results due to non-convergence of the MCMC chains. Therefore, we constrained the prior ranges to a shorter limit (as mentioned in Table 8) to get a reasonable fit. The results of the model fits obtained for both the hot Jupiters are listed in Table 8.

Similar to \cite{Mannaday20}, the derived values of the model parameters $e=0.00160^{+0.00096}_{-0.00105}$, $\omega_{0}=0.41^{+0.37}_{-0.29}$ rad and $\frac{d\omega}{dE}=0.000248^{+0.000144}_{-0.000155}$~rad~epoch$^{-1}$ are found to be statistically insignificant for TrES-3b. The reason behind the statistically insignificant results could be either due to a nearly circular orbit of TrES-3b, or strong correlation between model parameters, or considering the wrong model for the timing data. Except the less significant estimation of $\omega_{0}=0.23^{+0.25}_{-0.16}$ rad, the derived eccentricity of $e=0.00410^{+0.00062}_{-0.00110}$ and precession rate of $\frac{d\omega}{dE}=0.000440^{+0.000044}_{-0.000087}$~rad~epoch$^{-1}$ are found to be statistically significant for Qatar-1b. The eccentricity of Qatar-1b derived here appears to be compatible with the $1\sigma$ upper limit $(e\sim 0.012)$ reported in \cite{Bonomo17}.

Using the best-fit ephemeris derived for the apsidal precession and linear models, the timing residuals (i.e., ${T}^{c}_{ap}(E)$ - $T^{c}_{m}(E)$) were calculated and plotted as a function of epoch with blue dashed curve in the $O-C$ diagrams of TrES-3b and Qatar-1b (see Fig.~4). In this figure, the cyan solid lines are 100 random draws from the posteriors of the apsidal precession model, which are extrapolated for the next $\sim 4.5$ years to show the future trend of the apsidal precession phenomenon. The random draws of TrES-3b suggest that the apsidal precession model is consistent with the linear model, whereas those of Qatar-1b appear to be well representing the considered timing data. To confirm this, further high-precision photometric follow-up observations of the transits would be required.

\begin{table*}[ht]
\begin{center}
\caption {The Uniform Priors and Best-fit Model Parameters for TrES-3b and Qatar-1b}
\label{tab:5}
\small\addtolength{\tabcolsep}{-3pt}
\begin{tabular}{lllll}
\hline
\hline
{Parameter} & \multicolumn{2}{c}{Uniform Prior} & \multicolumn{2}{c}{Best-fit Values with $1\sigma$ Uncertainties}\\
& {TrES-3b} &	{Qatar-1b} & \multicolumn{1}{c}{TrES-3b} & \multicolumn{1}{c}{Qatar-1b}\\
\hline
{\bf Linear Ephemeris} & & &&\\
P [days]													& (0, 2) &	(0, 2)					&  ${1.30618628}^{+0.00000002}_{-0.00000002}$ & ${1.42002433}^{+0.00000003}_{-0.00000003}$\\
${T}_{0}$ [BJD$_{\rm TDB}-2,450,000$]							& (4184, 4186) &	(5646, 5649) 		& ${4185.911149}^{+0.000053}_{-0.000053}$ & ${5647.633175}^{+0.000050}_{-0.000050}$\\
\hline
\hline
$a_f$, $\tau$, $N_{\rm eff}$  									& 								& &$\sim 0.44, \sim 19, \sim 1052$  &$\sim 0.44, \sim 18, \sim 1111$ \\
$\chi^{2}$, $\chi^{2}_{red}(N_{\rm dof})^{a}$					& 							 	& &207.01, 1.63(127) &283.92, 1.56(182)\\
BIC														&								& &  216.73 & 294.35\\
\hline
{\bf Orbital Decay Ephemeris}& & &&\\
$P_q$ [days]												& (0, 2) & 	(0, 2)					& ${1.30618610}^{+0.00000012}_{-0.00000012}$  & ${1.42002386}^{+0.00000014}_{-0.00000014}$\\
${T}_{q0}$ [BJD$_{\rm TDB}-2,450,000$]									& (4184, 4186) & (5646, 5649) 			& ${4185.911241}^{+0.000079}_{-0.000079}$  &  ${5647.633357}^{+0.000073}_{-0.000073}$\\
$\frac{dP_q}{dE}$ [days]$^{b}$ 							& (-1, 1) &	(-1, 1)				& ${0.87899}^{+0.56241}_{-0.56517}$  & ${3.42245}^{ +1.01062}_{ -1.01495}$\\
\hline
\hline
$a_f$, $\tau$, $N_{\rm eff}$ 									&								& &$\sim 0.32, \sim 32, \sim 1000$& $\sim 0.32, \sim 35, \sim 914$ \\
$\chi^{2}$, $\chi^{2}_{red}(N_{\rm dof})$					& 								& & 204.12, 1.62(126)  &271.50, 1.50(181)\\
BIC															&								& & 218.70  & 287.14\\
\hline
{\bf Apsidal Precession Ephemeris} & & &&\\
$P_s$ [days]												& (0, 2) & (0, 2) 						& $1.30618619^{+0.00000008}_{-0.00000014}$ & ${1.42002379}^{+0.00000020}_{-0.00000020}$\\
$T_{ap0}$ [BJD$_{\rm TDB}-2,450,000$] 											& (4184, 4186) & (5646, 5649) 		& ${4185.911740}^{+0.000404}_{-0.000394}$  & ${5647.635063}^{+0.00039}_{-0.000543}$\\
e															& (0, 0.003) &	(0, 0.005)				& ${0.00160}^{+0.00096}_{-0.00105}$  & ${0.00410}^{+0.00062}_{-0.00110}$\\
$\omega_0$ [rad]											& (0, 1)	&	(0, 1)				& $0.41^{+0.37}_{-0.29}$  & $0.23^{+0.25}_{-0.16}$\\
$\frac{d\omega}{dE}$ [rad/Epoch]							& (0, 0.0005)&	(0, 0.0005)			& $0.000248^{+0.000144}_{-0.000155}$  & $0.000439^{+0.000044}_{-0.000087}$\\
\hline
\hline
$a_f$, $\tau$, $N_{\rm eff}$ 									& 								& &$\sim 0.35, \sim 197, \sim 1015$ & $\sim 0.41, \sim 166, \sim 1204$\\
$\chi^{2}$, $\chi^{2}_{red}(N_{\rm dof})$					& 								& & 207.08, 1.67(124) & 272.08, 1.52(179) \\
BIC															&								& & 231.38 & 298.15\\
\hline
\end{tabular} \\
{Note: $^a$ $N_{\rm dof}$ is the number of degree of freedom\\
$^{b}$ the uniform prior for $ \delta P$ is in days, while its best-fit value is in ${10}^{-10}$ days.
}
\end{center}
\end{table*}


\section{Discussion}
\subsection{Constraints on the Stellar Tidal Quality Factor}

Treating the observed $\frac{dP_q}{dE}$ of TrES-3b and Qatar-1b as a non-detection of orbital decay (see section 4.3), one can place a constraint on the tidal quality factor $Q^{'}_{\ast}$ for the host-stars of both hot Jupiters using the 5th percentile of the posterior probability distribution of $\frac{dP_q}{dE}$. To calculate ${Q}^{'}_{\ast}$, we used the following equation, which is the modified form of the constant-phase lag model of \citet{Goldr66} :
\begin{equation}
{Q}^{'}_{\ast} = -\frac{27}{2}{\pi}\left(\frac{M_p}{M_\ast}\right)\left(\frac{a}{R_\ast}\right)^{-5}\left(\frac{1}{\dot{P_q}}\right),
\end{equation}
where $P_q$ is the orbital period derived using the decay model, $\dot{P}_q$ is the period derivative, $\frac{M_p}{M_\ast}$ is the mass ratio of planet to star, and $\frac{a}{R_\ast}$ is the ratio of semi-major axis to stellar radius. For the TrES-3 system, the values of parameters $\frac{M_p}{M_\ast}=0.001964$ and $\frac{a}{R_\ast}=5.926$ are taken from \citet{Sozz09}. Corresponding to the 5th percentile of the posterior probability distribution of $\frac{dP_q}{dE}$, we derived ${\dot{P}_q}= -0.12$~ms~yr$^{-1}$ for TrES-3b. Substituting this $\dot{P}_q$ and the values of other parameters mentioned above in Equation (4), we constrain the tidal quality factor of TrES-3 to be ${Q}^{'}_{\ast}>2.78 \times {10}^{6}$ with 95\% confidence \citep{Maci18,Patra20}. As the 5th percentile value of the posterior probability distribution of $\frac{dP_q}{dE}$ obtained for Qatar-1b is positive,  we cannot place any constraint on ${Q}^{'}_{\ast}$ for Qatar-1.

\subsection{Estimations of Planetary Love Number ($k_{p}$)}

If we assume that the observed TTVs in both the hot Jupiter systems are caused by apsidal precession, then the respective orbital eccentricity of TrES-3b and Qatar-1b would be ${0.00160}^{+0.00096}_{-0.00105}$ and ${0.00410}^{+0.00062}_{-0.00110}$ (see Section 4.4). According to tidal evolution theory, the orbits of the hot Jupiters are expected to be circularized on a shorter timescale than their system's ages \citep{Levrd07,Dawson18}. Using Equation (17) of \cite{Patra17} with an assumption of planetary tidal dissipation, ${Q}^{'}_{p}=10^6$, the estimated timescales for tidal circularization of the orbits of TrES-3b and Qatar-1b are $\sim 4$ Myr and $\sim 7$ Myr, respectively. These timescales are several orders of magnitude smaller than the ages of their host stars (TrES-3: $0.6^{+2.0}_{-0.4}$ Gyr; \citealt{Matsu10}, Qatar-1: $\sim 4$ Gyr; \citealt{Covino13}). As the timescales for tidal circularization are much smaller than the ages of the systems, the presence of an orbital eccentricity is inconsistent with tidal theory. If apsidal precession is taking place in these systems, there must be some mechanism that can excite and maintain a non-zero eccentricity \citep[see][]{Bouma19,Maci21}. In this context, \citet{Ragoz09} showed that the interiors of very hot Jupiters could be a dominant source for apsidal precession. Since the apsidal precession rate is proportional to planetary tidal Love number, $k_p$, a dimensionless parameter that depends on the interior density distribution of the planet, we adopted the following equation of \citet{Patra17}:
\begin{equation}
\frac{d\omega}{dE}={15}{\pi}{k_p}\left(\frac{M_\ast}{M_p}\right)\left(\frac{R_p}{a}\right)^5,
\end{equation}
and estimated the value of $k_p$ for both the hot Jupiters. By substituting the values of $\frac{d\omega}{dE}$ listed in Table 8, and other relevant parameters taken from \citet{Sozz09} and \citet{Maci16} in the above Equation (5), we found $k_{p}=0.61 \pm 0.12$ and $k_{p}=1.95 \pm 0.26$ for TrES-3b and  Qatar-1b, respectively. From the estimated $k_p =0.61 \pm 0.12$, the interior density distribution of TrES-3b appears to be similar to that of Jupiter \citep[$k_p=0.59:$][]{Wahl16}. As the obtained value of $k_{p}= 1.95 \pm 0.26$ for Qatar-1b is an unphysical value with a large uncertainty, it is difficult to infer the interior density profile of this hot Jupiter. This indicates a non-detection of apsidal precession using the available timing data of Qatar-1b \citep{Maci21}. To confirm this, future follow-up observations of transits and occultations are needed.

\subsection{Applegate Mechanism}

It was proposed by \citet{Watson10} that changes in the quadruple  moment of the host star \citep{Applegate92} may also induce long-term TTVs in exoplanetary systems. Considering the modulation periods of 11, 22 and 50 yr for magnetic activity cycles, they calculated the possible TTV amplitudes for many stars including TrES-3. The largest TTV amplitude expected by them for TrES-3b due to Applegate mechanism was $\delta{t} \sim 3.1$~s, which is an order of magnitude smaller than the TTV amplitude of $\sim 32.26$ s found by us (see Section 4.1). According to Equation (13) of \citet{Watson10}, the largest TTV amplitude that can be induced in Qatar-1 system due to this mechanism in a modulation period of 50 yr would be $\delta{t} \le 3$ s, which is much smaller than the TTV amplitude of $\sim 113.65$ s calculated by us (see Section 4.1). Therefore, an Applegate mechanism may not be a possible cause of TTVs in the TrES-3 and Qatar-1 systems.

\subsection{Line-of-sight Acceleration}

An accelerating motion of the center of mass of the star-planet system toward or away from the observer's line-of-sight can also change the observed orbital period of an extrasolar planet (see WASP-4b: \citealt{Bouma19,Bouma20,South19,Turner22}; TrES-5b: \citealt{Maci21}). In contrast to orbital decay, the sign of the period change could either be positive or negative, depending on the direction of acceleration. Acceleration toward the observer causes a decreasing period, whereas an increasing period corresponds to acceleration away from the observer. If the center of mass of the TrES-3 and Qatar-1 systems were really accelerating away from us with an amplitude ${\dot{v}}_{RV}$, then the observed $\dot{P_q}$ of their hot Jupiters would be related to ${\dot{v}}_{RV}$ by the following equation of \citet{Maci21}:

\begin{equation}
\dot{v}_{RV}=\frac{\dot{P_q}}{P_q} {c},
\end{equation}

where $c$ is the speed of light and ${\dot{v}}_{RV}$ is the linear trend in the RV of the host star. By substituting the derived values of $P_q$ and $\dot{P_q}$ (see Section 4.3) in Equation (6), the values of ${\dot{v}}_{RV}$  are found to be $\sim 0.015 \pm 0.014$ ~m~s$^{-1}$~d$^{-1}$ and $ \sim 0.051 \pm 0.021$ m s$^{-1}$ d$^{-1}$ for TrES-3 and Qatar-1, respectively. In order to confirm this, the RV data of both the systems were collected from the literature (TrES-3: \citealt{Sozz09,Knutson14}; Qatar-1: \citealt{Covino13,Bonomo17}) and modeled using the $radvel$ package \citep{Fulton18}. The parameters RV semi-amplitude ($K_b$), zero-point velocity ($\gamma$), linear trend in RV ($\dot{\gamma}={{\dot{v}}_{RV}}$) and jitter term ($\sigma$) were fitted freely. The orbital period and mid-transit time were fitted under a Gaussian penalty with their $1\sigma$ uncertainties from \citet{Mannaday20} and \citet{Su21}. The remaining parameters such as eccentricity ($e$), argument of periastron ($\omega$), and quadratic trend in RV ($\ddot{\gamma}$) were fixed to zero. By analyzing the RV data of TrES-3, we find $\dot{\gamma} = 0.157 \pm 0.056$~m~s$^{-1}$~d$^{-1}$, which is 10 times larger than the above mentioned value of ${\dot{v}}_{RV}$. From this, it appears that the observed TTV in the TrES-3 system may not be due to the line-of-sight acceleration. Since the value of $\dot{\gamma} = 0.046 \pm 0.008$~m~s$^{-1}$~d$^{-1}$ obtained from RV data analysis of Qatar-1 is found to be consistent within 1$\sigma$ limit of the above estimated value of ${\dot{v}}_{RV}$, an acceleration of Qatar-1 away from our line-of-sight is a possible cause of the observed TTV. In order to confirm this, future high precision RV observations of Qatar-1 are required.

\subsection{Possible Explanation for the Observed TTVs}

For TrES-3b, the difference in the BIC between the apsidal precession and linear models is $\bigtriangleup\mathrm{BIC} = \mathrm{BIC}_{precession} - \mathrm{BIC}_{linear} =14.65$ with an approximate Bayes factor of $\exp(\bigtriangleup\mathrm{BIC}/2) = 1517.77$. This $\bigtriangleup\mathrm{BIC}$ provides strong evidence to rule out the possibility of apsidal precession in this planetary system \citep[cf.][]{Blecc14,Maci16,Patra17,Mannaday20,Athano22}. This is also justified by the fact that the estimated value of the eccentricity, $e={0.00160}^{+0.00096}_{-0.00105}$, is very small, which would have a marginal effect on the mid-transit times determined from the transit light curves \citep{Maci16,Mannaday20}. As the $\bigtriangleup\mathrm{BIC}=\mathrm{BIC}_{decay}-\mathrm{BIC}_{linear}=1.97$ is found to be smaller than 2, it is difficult to clearly distinguish between the orbital decay and linear model \citep{Colln17}. However, the positive and statistically less significant estimated value of $\dot{P_q}$ (see Section 4.3), as well as non-detection of the Applegate mechanism and line-of-sight acceleration in the TrES-3 system enabled us to prefer the linear model to explain the transit time data of TrES-3b.

Based on the smaller values of ${\chi}^{2}_{red}$ and BIC obtained in the orbital decay model fit as compared to other models fit (see Table 8), the orbital decay model appears to be the best model for the presently considered transit time data of Qatar-1b. However, we do not prefer this model because the estimated positive value of $\dot{P_q}$ indicates an increasing period. Due to having $\bigtriangleup\mathrm{BIC}=\mathrm{BIC}_{precession} - \mathrm{BIC}_{linear}=3.84$ with an approximate Bayes factor of $\exp(\bigtriangleup\mathrm{BIC}/2)= 6.82$, the linear model is somewhat preferable over the apsidal precession model for the considered transit time data of Qatar-1b. In addition to this, the unphysical value of $k_p$ inferred from the observed precession rate (see Section 5.2) does not favor the presence of apsidal precession in the Qatar-1 system. From the RV data analysis of Qatar-1, it appears that the observed TTV of its hot Jupiter is caused due to line-of-sight acceleration of the system away from the Sun. However, future transit and RV observations of Qatar-1 would be valuable to clearly distinguish between constant and increasing period phenomena.


\section{Concluding Remarks}

We present all the transit light curves of TrES-3b and Qatar-1b observed by TESS in sectors 25, 26, 40, and sectors 17, 21, 24, 25, 41, 48, respectively. In addition to the TESS light curves, two more transits of Qatar-1b observed by us using a 1.23 m telescope are presented. Since our aim was to refine the transit ephemeris and examine the possibility of TTVs in both systems, we have also collected the best-quality light curves from the ETD and the literature. In total 182 transit light curves of TrES-3b and 228 transit light curves of Qatar-1b spanning over more than a decade are employed in this work. Fitting a linear ephemeris model to the precisely estimated transit timing data, we have obtained refined orbital ephemerides for both systems. The derived ephemerides are fully consistent with previous results but are estimated with improved precision (see Table 8). Our timing analysis indicates the possibility of TTVs in both hot Jupiters, which are unlikely to be a short-term and periodic. This enables us to rule out the presence of additional planets in orbits close to TrES-3b and Qatar-1b.

Motivated by the above results, we have examined the possibilities of orbital decay and apsidal precession phenonmena in both hot Jupiters. For TrES-3b and Qatar-1b, the respective positive period derivatives of $\dot{P_q} =2.12 \pm 1.92$~ms~yr$^{-1}$ and $\dot{P_q}=7.60 \pm 3.19$~ms~yr$^{-1}$ indicate non-detection of orbital decay in both the systems. However, we were able to constrain the lower limit of the tidal quality factor of TrES-3 to be ${Q}^{'}_{\ast}>\ 2.78 \times {10}^{6}$ with 95\% confidence. We were not able to place a constraint on ${Q}^{'}_{\ast}$ for Qatar-1 because the 95\% lower limit on $\dot{P_q}$ is still positive. For TrES-3b, the value of ${k}_p$ calculated with its observed precession rate of $\frac{d\omega}{dE}=0.000248^{+0.000144}_{-0.000155}$~rad~epoch$^{-1}$ is found to be $0.61 \pm 0.12$. This value of $k_p$ suggests that the interior density distribution of TrES-3b appears to be similar to that of Jupiter. Because of the statistically less significant estimated value of $\frac{d\omega}{dE}$, further high precision transit observations are required to confirm the above inferred interior density profile of TrES-3b. Although the observed precession rate of Qatar-1b is statistically significant, the interior density profile of this hot Jupiter cannot be specified because the inferred value of $k_p$ is unphysical.

The Applegate mechanism is not a possible cause of the observed TTVs in both systems, since the largest TTV amplitudes expected from this mechanism are an order of magnitude smaller than the observed TTV amplitudes. Moreover, the observed TTV of TrES-3b does not appear to originate from line-of-sight acceleration as the value of $\dot{\gamma}$ derived from RV data is 10 times larger than $\dot{v}_{RV}$ estimated using the observed $\dot{P_q}$. In contrast to TrES-3, the line-of-sight acceleration of Qatar-1 is a possible explanation for the observed period change. In order to confirm our findings, we propose further photometric follow-up observations of the primary and secondary eclipse, as well as RV measurements of these hot Jupiter systems. In this regard, it is worth mentioning that the ongoing TESS observations of TrES-3 (sectors 52-53) and Qatar-1 (sector 51 and sectors 55-59) systems would be useful to shed some more lights on the results obtained in this paper.


\section*{Acknowledgments}

We thank the anonymous referee for useful comments that improved the quality of the paper. VKM and PT thank the staff at IAO, Hanle and CREST (IIA), Hosakote for providing support during the previous observations of Qatar-1b and TrES-3b. PT expresses his sincere thanks to IUCAA, Pune for providing the supports through IUCAA Associateship Programme. PT and VKM acknowledge the University Grants Commission (UGC), New Delhi for providing the financial support through Major Research Project no. UGC-MRP 43-521/2014(SR). JS and LM thank the staff at CAHA, Spain, for the observational supports. IGJ acknowledges funding from the Ministry of Science and Technology, Taiwan, through the grant No.\ MOST 110-2112-M-007-035. This work was supported by the Slovak Research and Development Agency under contract No. APVV-20-0148. This work was also supported by a VEGA grant of the Slovak Academy of Sciences, grant No. 2/0031/22. The research of PG was supported by an internal grant VVGS-PF-2021-2087 of the Faculty of Science, P.J. \v{S}af\'{a}rik University in Ko\v{s}ice. This paper includes data collected by the TESS mission, which are publicly available in the Mikulski Archive for Space Telescopes (MAST). The contributor to Exoplanet Transit Database (ETD) are gratefully acknowledged for making their transit light curves publicly available. While performing timing analysis the fruitful discussions and valuable suggestions received from D. Ragozzine and Kishore C. Patra are also acknowledged. We thank N.\ P.\ Gibson, J.\ W.\ Lee, D.\ Ricci, D.\ Mislis and K.\ A.\ Collins for sharing their light curves with us.  We also thank A.\ Sozzetti, K.\ D.\ Col\'{o}n, P.\ Kundurthy, J.\ D.\ Turner, E.\ Covino, C.\ von Essen, and G.\ Maciejewski for making their transit light curves publicly available. This paper has made use of the VizieR catalogue access tool, operated at CDS, Strasbourg, France, and NASA’s Astrophysics Data System Bibliographic Services.

\software: juliet \citep{Espinoza19}, celerite \citep{Foreman17}, MultiNest \citep{Feroz09}, PyMultinest \citep{Buchner14}, TAP \citep{Gaz12}, JKTLD \citep{South15},  emcee \citep{Foreman13}, Radvel \citep{Fulton18}, {IP}ython \citep{Perez07}, corner \citep{Foreman16}, astroML \citep{VanderPlas12}, NumPy \citep{vander11}, and SciPy \citep{Jones01}.

\begin{appendix}
\section{TESS observed transit light curves of TrES-3b and Qatar-1b}
 \setcounter{figure}{0}
\renewcommand{\thefigure}{A\arabic{figure}}
 \begin{figure*}[ht]
	\includegraphics[width=\columnwidth]{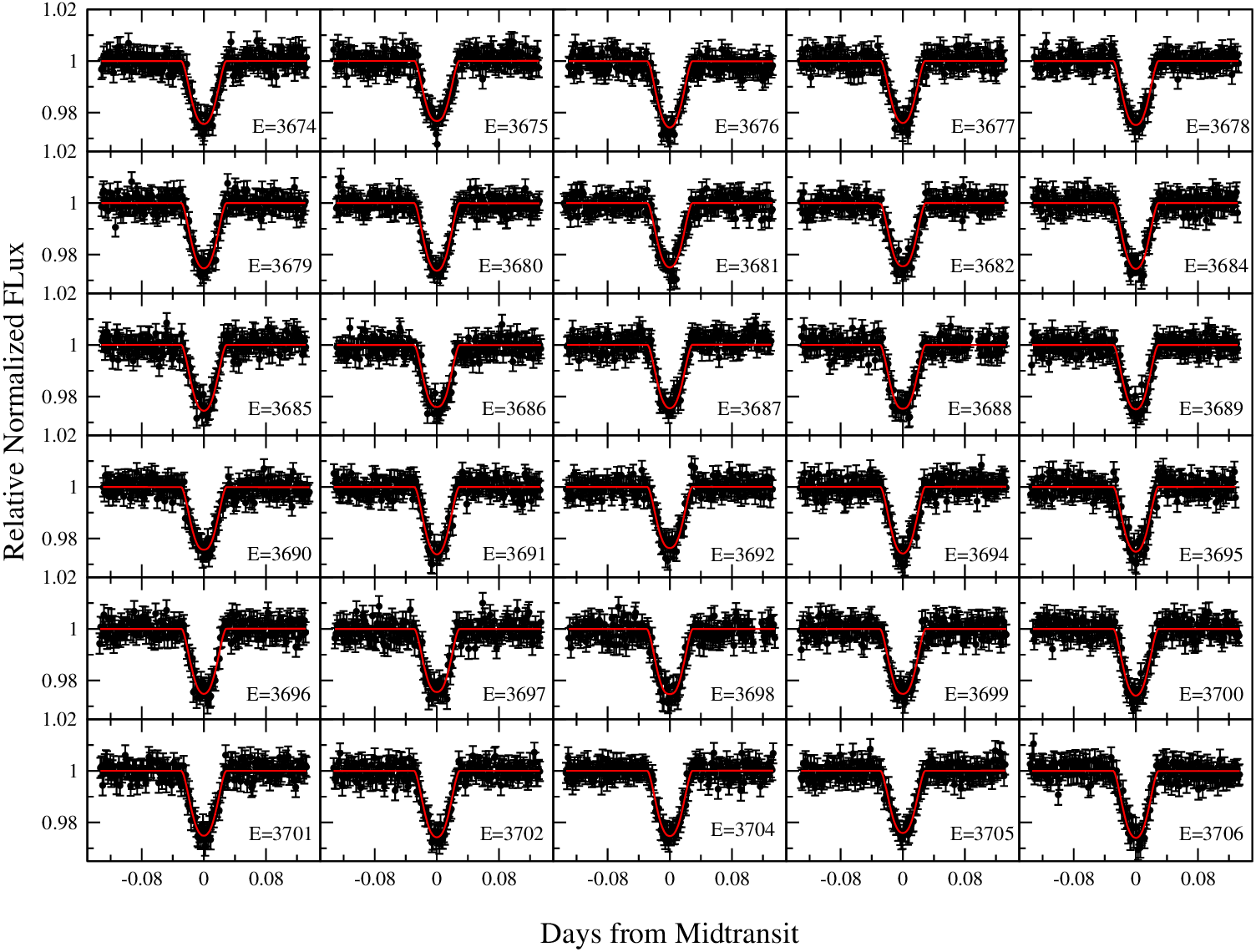}
    \caption{The normalized relative flux of TrES-3 as a function of the time (the offset from mid-transit time and in TDB-based BJD) of individual transit observed by TESS: points are the data, solid lines are best-fit models and E is the calculated epoch number.}
    \label{fig:5}
\end{figure*}

\begin{figure*}[ht]
	\includegraphics[width=\columnwidth]{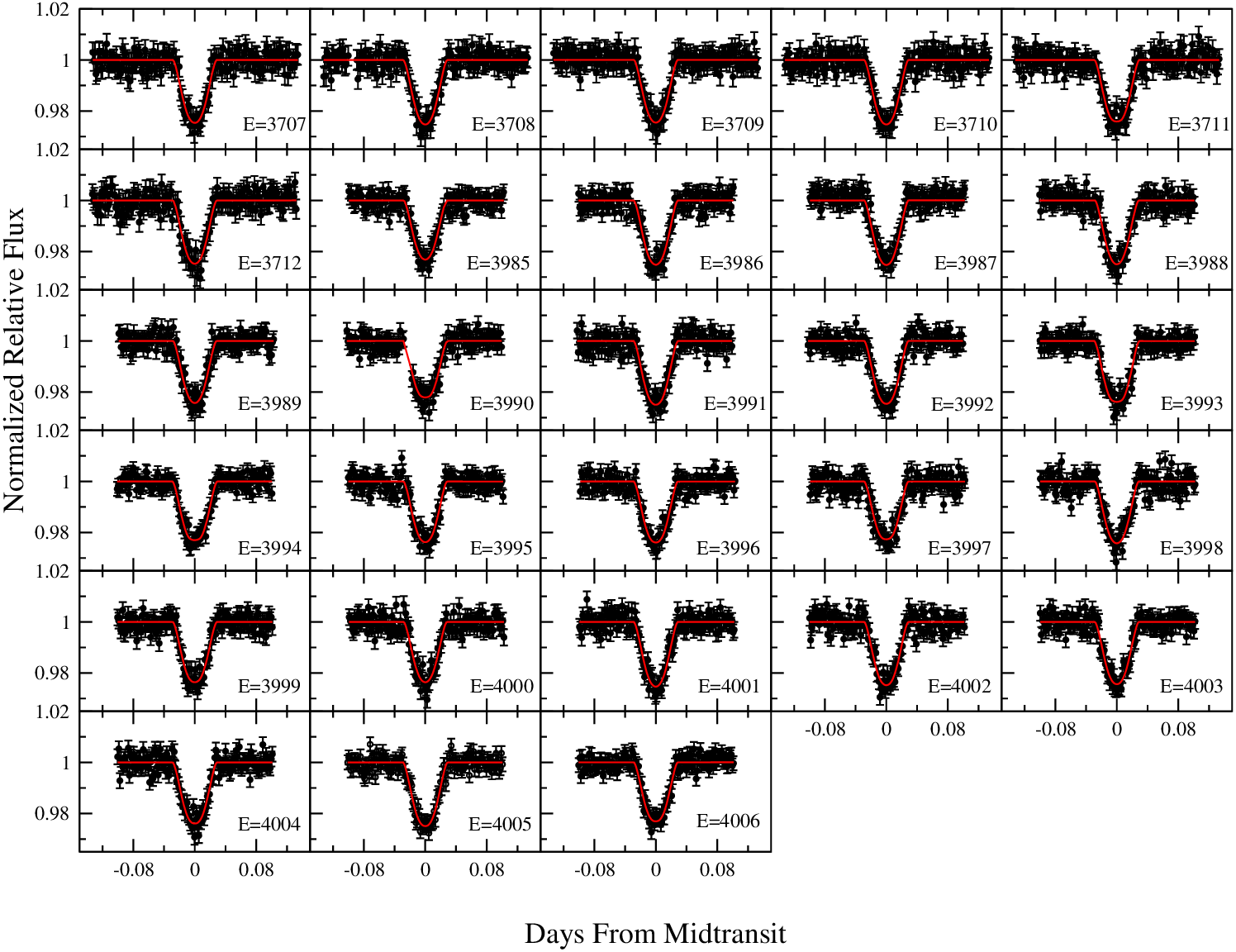}
    \caption{Same as the Figure A1 but for remaining epochs.}
    \label{fig:6}
\end{figure*}

\begin{figure*}[ht]
	\includegraphics[width=\columnwidth]{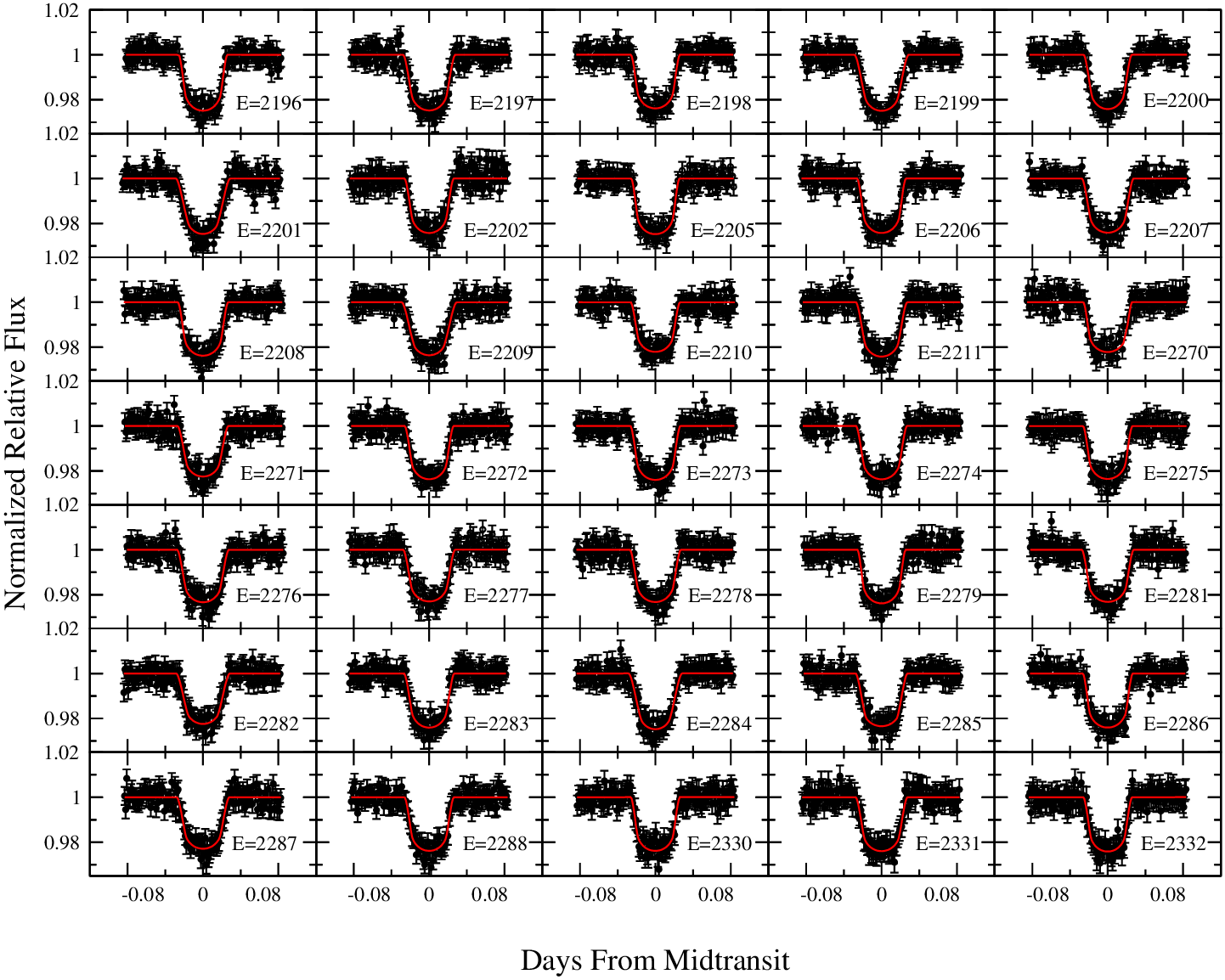}
    \caption{The normalized relative flux of Qatar-1 as a function of the time (the offset from mid-transit time and in TDB-based BJD) of individual transit observed by TESS: points are the data, solid lines are best-fit models and E is the calculated epoch number.}
    \label{fig:7}
\end{figure*}

\begin{figure*}[ht]
	\includegraphics[width=\columnwidth]{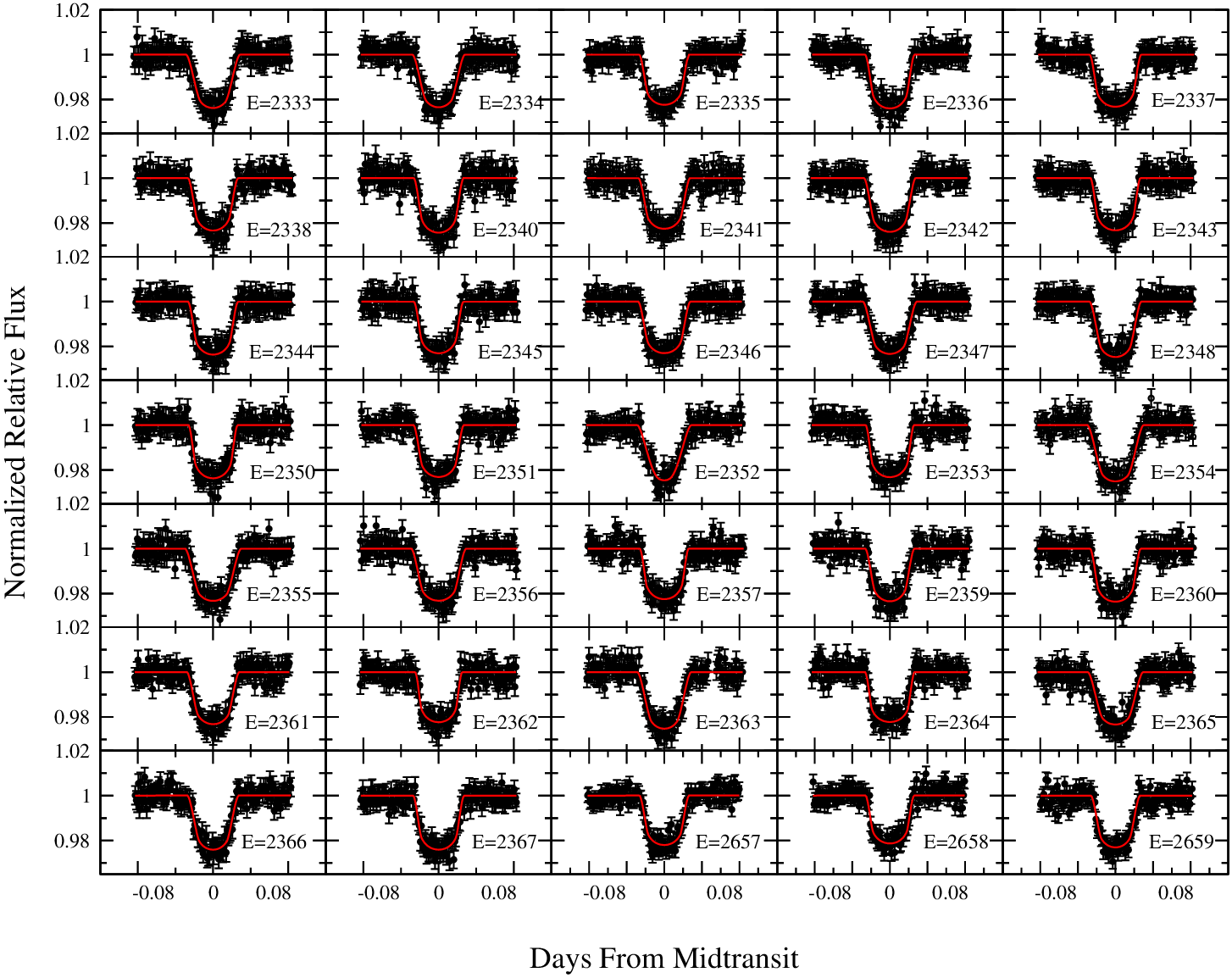}
    \caption{Same as the Figure A3 but for remaining epochs.}
    \label{fig:8}
\end{figure*}

\begin{figure*}[ht]
	\includegraphics[width=\columnwidth]{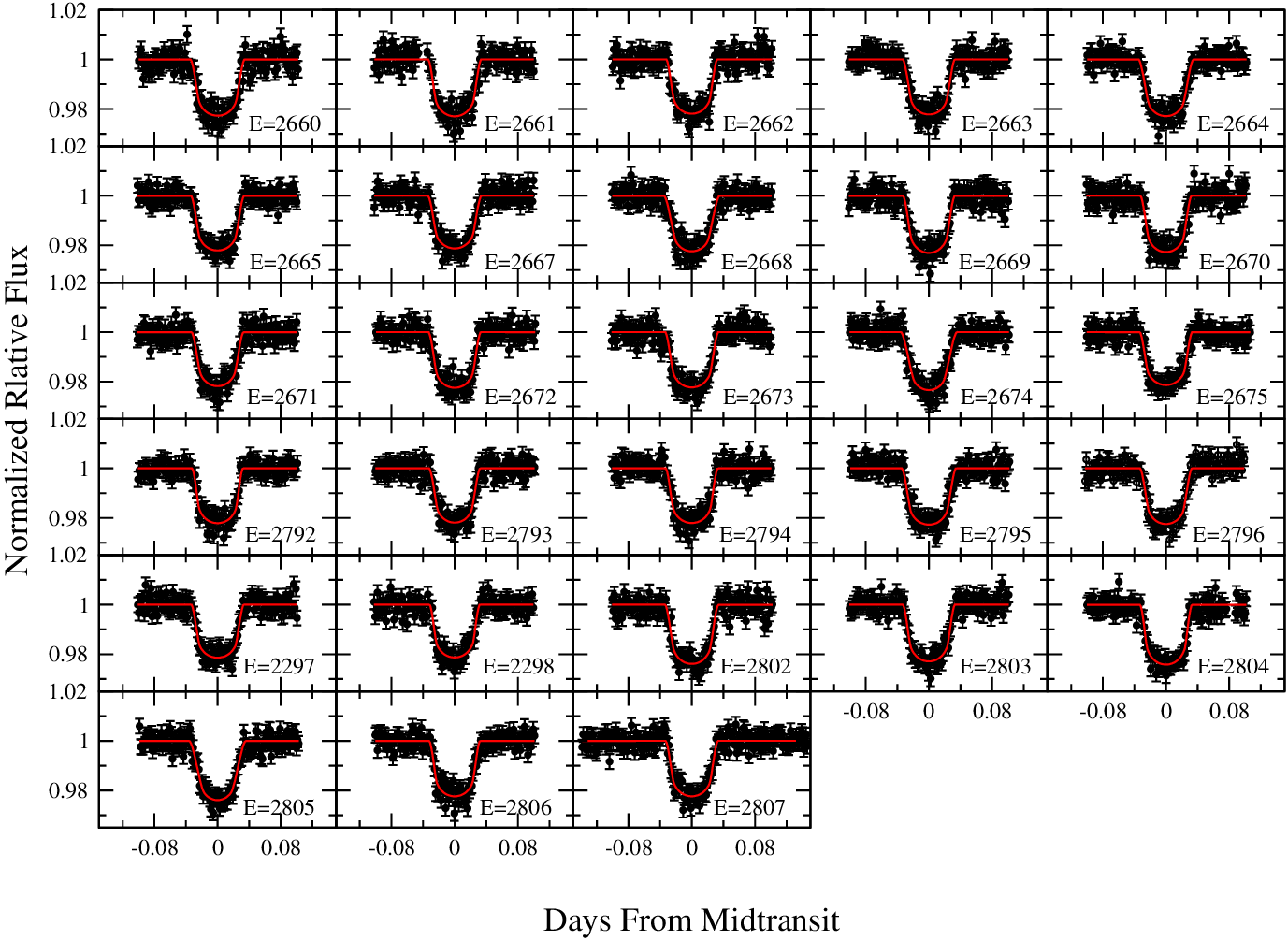}
    \caption{Same as the Figure A3 but for remaining epochs.}
    \label{fig:9}
\end{figure*}

\end{appendix}



\end{document}